\pdfoutput=1
\documentclass[a4paper,cits]{JINST}
\usepackage{microtype}
\usepackage{booktabs}
\usepackage{amsmath}
\usepackage{lineno}
\usepackage{enumerate}

\let\ifpdf\relax

\newcommand{\bbonu}{\ensuremath{\beta\beta0\nu}}
\newcommand{\bbtnu}{\ensuremath{\beta\beta2\nu}}
\newcommand{\Qbb}{\ensuremath{Q_{\beta\beta}}}

\newcommand{\NA}{\ensuremath{^{22}}Na}

\newcommand{\XE}{\ensuremath{{}^{136}\rm Xe}}

\author{
\mbox{The NEXT Collaboration}

D.~Lorca,$^{a}$\thanks{Co-corresponding author} ~
J.~Mart\'in-Albo,$^{a}$\thanks{Co-corresponding author} ~
A.~Laing,$^{a}$
P.~Ferrario,$^{a}$
J.J.~G\'omez-Cadenas,$^{a}$\thanks{Spokesperson (gomez@mail.cern.ch)} 

V.~\'Alvarez,$^{a}$
F.I.G.~Borges,$^{b}$
M.~Camargo,$^{h}$
S.~C\'arcel,$^{a}$
S.~Cebri\'an,$^{c}$
A.~Cervera,$^{a}$
C.A.N.~Conde,$^{b}$
T.~Dafni,$^{c}$
J.~D\'iaz,$^{a}$
R.~Esteve,$^{e}$
L.M.P.~Fernandes,$^{b}$
A.L.~Ferreira,$^{g}$
E.D.C.~Freitas,$^{b}$
V.M.~Gehman,$^{d}$
A.~Goldschmidt,$^{d}$
H.~G\'omez,$^{c}$
D.~Gonz\'alez-D\'iaz,$^{c}$
R.M.~Guti\'errez,$^{h}$
J.~Hauptman,$^{i}$
J.A.~Hernando Morata,$^{j}$
D.C.~Herrera,$^{c}$
I.G.~Irastorza,$^{c}$
L.~Labarga,$^{k}$
I.~Liubarsky,$^{a}$
M.~Losada,$^{h}$
G.~Luz\'on,$^{c}$
A.~Mar\'i,$^{e}$
G.~Mart\'inez-Lema,$^{j}$
A.~Mart\'inez,$^{a}$
T.~Miller,$^{d}$
F.~Monrabal,$^{a}$
M.~Monserrate,$^{a}$
C.M.B.~Monteiro,$^{b}$
F.J.~Mora,$^{e}$
L.M. Moutinho,$^{g}$
J.~Mu\~noz~Vidal,$^{a}$
M.~Nebot-Guinot,$^{a}$
D.~Nygren,$^{d}$
C.A.B.~Oliveira,$^{d}$
J.~P\'erez,$^{k}$
J.L.~P\'erez~Aparicio,$^{l}$
J.~Renner,$^{d}$
L.~Ripoll,$^{m}$
A.~Rodr\'iguez,$^{c}$
J.~Rodr\'iguez,$^{a}$
F.P.~Santos,$^{b}$
J.M.F.~dos~Santos,$^{b}$
L.~Segu\'i,$^{c}$
L.~Serra,$^{a}$
D.~Shuman,$^{d}$
A.~Sim\'on,$^{a}$
C.~Sofka,$^{n}$
M.~Sorel,$^{a}$
J.F.~Toledo,$^{d}$
J.~Torrent,$^{m}$
Z.~Tsamalaidze,$^{f}$
J.F.C.A.~Veloso,$^{g}$
R.~Webb,$^{n}$
J.T.~White,$^{n}$\thanks{deceased} ~
N.~Yahlali$^{a}$\\
\llap{$^{a}$}
Instituto de F\'isica Corpuscular (IFIC), CSIC \& Universitat de Val\`encia\\
Calle Catedr\'atico Jos\'e Beltr\'an, 2, 46980 Paterna, Valencia, Spain\\
\llap{$^{b}$}
Departamento de Fisica, Universidade de Coimbra\\
Rua Larga, 3004-516 Coimbra, Portugal\\
\llap{$^c$}
Lab.\ de F\'isica Nuclear y Astropart\'iculas, Universidad de Zaragoza\\ 
Calle Pedro Cerbuna, 12, 50009 Zaragoza, Spain\\
\llap{$^d$}
Lawrence Berkeley National Laboratory (LBNL)\\
1 Cyclotron Road, Berkeley, California 94720, USA\\
\llap{$^{e}$}
Instituto de Instrumentaci\'on para Imagen Molecular (I3M), Universitat Polit\`ecnica de Val\`encia\\ 
Camino de Vera, s/n, Edificio 8B, 46022 Valencia, Spain\\
\llap{$^{f}$}
Joint Institute for Nuclear Research (JINR)\\
Joliot-Curie 6, 141980 Dubna, Russia\\
\llap{$^{g}$}Institute of Nanostructures, Nanomodelling and Nanofabrication (i3N), Universidade de Aveiro\\
Campus de Santiago, 3810-193 Aveiro, Portugal\\
\llap{$^{h}$}
Centro de Investigaciones, Universidad Antonio Nari\~no\\ 
Carretera 3 este No.\ 47A-15, Bogot\'a, Colombia\\
\llap{$^{i}$}
Department of Physics and Astronomy, Iowa State University\\
12 Physics Hall, Ames, Iowa 50011-3160, USA\\
\llap{$^{j}$}
Instituto Gallego de F\'isica de Altas Energ\'ias (IGFAE), Univ.\ de Santiago de Compostela\\
Campus sur, R\'ua Xos\'e Mar\'ia Su\'arez N\'u\~nez, s/n, 15782 Santiago de Compostela, Spain\\
\llap{$^{k}$}
Departamento de F\'isica Te\'orica, Universidad Aut\'onoma de Madrid\\
Ciudad Universitaria de Cantoblanco, 28049 Madrid, Spain\\
\llap{$^{l}$}
Dpto.\ de Mec\'anica de Medios Continuos y Teor\'ia de Estructuras, Univ.\ Polit\`ecnica de Val\`encia\\
Camino de Vera, s/n, 46071 Valencia, Spain\\
\llap{$^{m}$}
Escola Polit\`ecnica Superior, Universitat de Girona\\
Av.~Montilivi, s/n, 17071 Girona, Spain\\
\llap{$^{n}$}
Department of Physics and Astronomy, Texas A\&M University\\
College Station, Texas 77843-4242, USA\\

E-mail: \email{david.lorca@ific.uv.es, justo.martin-albo@ific.uv.es}
}

\title{Characterisation of NEXT-DEMO using xenon K$\mathbf{_\alpha}$ X-rays}

\abstract{The NEXT experiment aims to observe the neutrinoless double
beta decay of \XE\ in a high-pressure xenon gas TPC using
electroluminescence (EL) to amplify the signal from
ionization. Understanding the response of the detector is imperative
in achieving a consistent and well understood energy measurement. The
abundance of xenon K-shell X-ray emission during data taking has been
identified as a multitool for the characterisation of the fundamental
parameters of the gas as well as the equalisation of the response of
the detector.

The NEXT-DEMO prototype is a $\sim$1.5 kg volume TPC filled with
natural xenon. It employs an array of 19 PMTs as an energy plane and
of 256 SiPMs as a tracking plane with the TPC light tube and SiPM
surfaces being coated with tetraphenyl butadiene (TPB) which acts as a
wavelength shifter for the VUV scintillation light produced by
xenon. This paper presents the measurement of the properties of the
drift of electrons in the TPC, the effects of the EL production
region, and the extraction of position dependent correction
constants using $K_\alpha$ X-ray deposits. These constants were
used to equalise the response of the detector to deposits left by
gammas from \NA.}

\keywords{Time projection chambers (TPC); Gaseous imaging and tracking detectors; SiPM; Xenon properties}

\begin{document}

\tableofcontents

\newpage
\section{Introduction} \label{sec:Introduction}
Gas detector and, particularly, time projection chamber (TPC) technology have been developed over a number of years and now find applications in a diverse range of fields. The relative ease with which noble gases can be cleaned of impurities as well as the availability of both scintillation and ionization signals when using these materials as target has meant that such detectors have been extensively used in medical imaging, dark matter detection, X-ray astronomy, and for the observation of double beta decay (see for example \cite{Iacobaeus:2004im,Akerib:2012ys,Ohashi:1996zf,Alvarez:2012haa}).

Xenon, both as a gas and liquid, is of special interest for applications in which the energy of the interacting particle must be measured. In particular, gas Xenon presents a low Fano factor, which results in a good intrinsic limit on the energy resolution, 0.3\% FWHM at 2500~keV \cite{Nygren:2009NIM}. This trait coupled with the existence of the \XE\ isotope which can decay via the double beta mechanism (\bbtnu) makes xenon an attractive material for the search for neutrinoless double beta decay (\bbonu). Moreover, the long lifetime of the 2 neutrino mode, $2.165\times 10^{21}$~yr \cite{PhysRevC.89.015502}, is advantageous due to the reduction factor afforded in terms of the background from this mechanism.

NEXT will use a target mass of high purity \XE\ gas at high pressure to search for the double electron events indicative of double beta decays. The scintillation light produced by the interaction will be used as the start of event with the ionization electrons being induced to produce more light after drift in a region of higher electric field via the process of Electroluminescence (EL). This light will be detected in PMTs at the opposite end of the drift region where a measurement of the event energy will be made, and in an array of silicon photomultipliers. This tracking plane, positioned $\sim$2~mm behind the EL region will be used to perform a topological reconstruction and pattern recognition of the events.

\begin{figure}[b]
\centering
\includegraphics[width=0.90\textwidth]{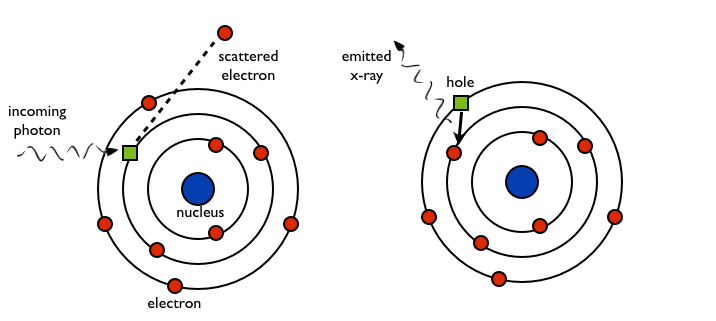}
\caption{X-ray emission process. An incoming photon with enough energy extracts an electron from the K, L shells (left). The hole is filled by a more energetic electron with the consequent emission of an X-ray (right).} \label{fig:xrayprocess}
\end{figure}

In xenon, as in most noble gases, there exists a non-zero probability that an interacting photon will excite the K or L shells of an atom causing the emission of an X-ray. This can take place in two ways (illustrated in figure~\ref{fig:xrayprocess}):
\begin{enumerate}[1-]

\item The gamma directly excites an electron which upon de-excitation emits a K/L-shell X-ray.
\item An electron from the K/L shells is knocked out of the xenon atom. The hole is then filled by an electron from a higher shell, which emits an X-ray.

\end{enumerate}

\begin{figure}
\centering
\includegraphics[width=0.8\textwidth]{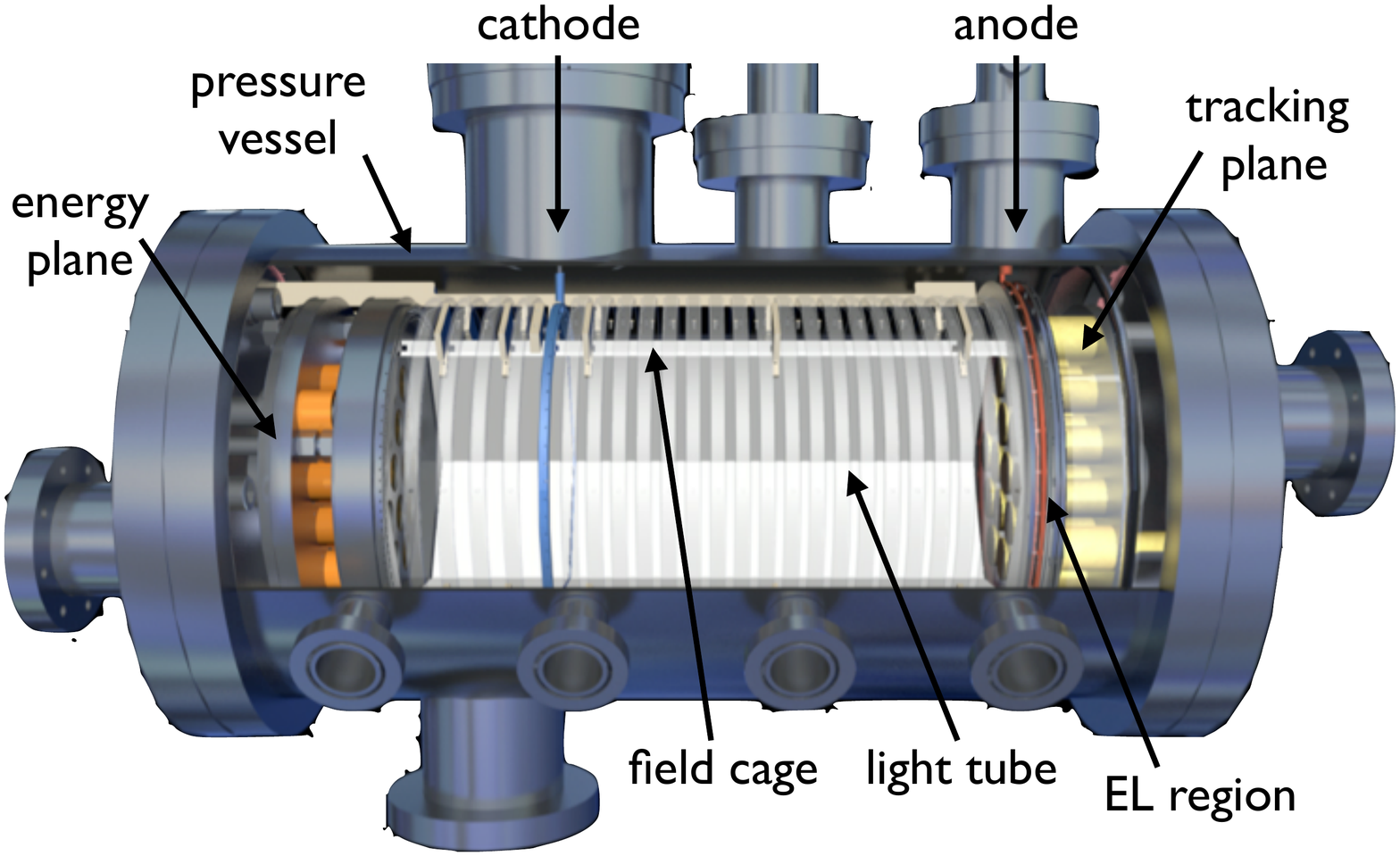}
\caption{Schematic of NEXT-DEMO.} \label{fig:NEXT-DEMO}
\end{figure}

In both cases, at 10 bar the X-ray will travel on average $\sim$1.39~cm \cite{NISTTABLE} before interacting with the gas, producing a photoelectric electron. In xenon in the range of sensitivity of NEXT the most important lines are the $K_\alpha$ and $K_\beta$ emissions with 29.7~keV and 33.8~keV respectively. Electrons produced at these energies will travel a maximum CSDA distance $\sim$0.6~mm at 10~bar \cite{NISTESTAR} but will tend to displace from their production point by less due to the dominance of multiple scattering. The abundance of such events and their effectively point-like nature make K-shell X-ray interactions useful tools for the calibration of the detector. They can be used to study fundamental properties of the gas and drift region as well as to equalise the energetic response which varies due to detector geometry. The NEXT-DEMO prototype \cite{Alvarez:2012xda,Alvarez:2013gxa}, shown in figure \ref{fig:NEXT-DEMO}, has been used to study these properties.

\section{Experimental setup} \label{sec:ExpSet}
NEXT-DEMO is a cylindrical pressure vessel made of stainless steel, able to withstand up to 20~bar of internal pressure. It is $60$~cm long and $30$~cm in diameter, and holds $\sim1.5$ kg of Xe at 10~bar. Three wire grids, the \emph{cathode}, \emph{gate} and \emph{anode}, limit the two active regions of the TPC. When a charged particle or photon interacts with the xenon, both scintillation light and electron-ion pairs are produced. The prompt scintillation light is emitted in the VUV ($\sim 178$ nm) and directly detected in a plane of 19 Hamamatsu R7378A PMTs (the \emph{energy plane}) behind the cathode grid, this light (known as S1) defines the start of an event and is generally used as an event read-out trigger. The ionization electrons are induced to drift towards the gate by an electric field (generally of 500~V~cm$^{-1}$ except in the data of section \ref{sec:PropXTPC}, where the field was varied within 200 - 500~V~cm$^{-1}$) where they enter in a region of higher field and are induced to produce further scintillation light through the process of electroluminescence (EL). This secondary light is once again detected in the energy plane but the forward going photons are also detected in an array of 256 tetraphenyl butadiene (TPB) coated Hamamatsu S10362-11-050P SiPMs. This \emph{tracking plane} is used to reconstruct the position of energy deposits and, ultimately, the topology of an event as a whole.

\subsection{Calibration of Sensors} 
\label{subsec:CalSen}
\begin{figure}
  \begin{center}
    \includegraphics[width=0.49\textwidth]{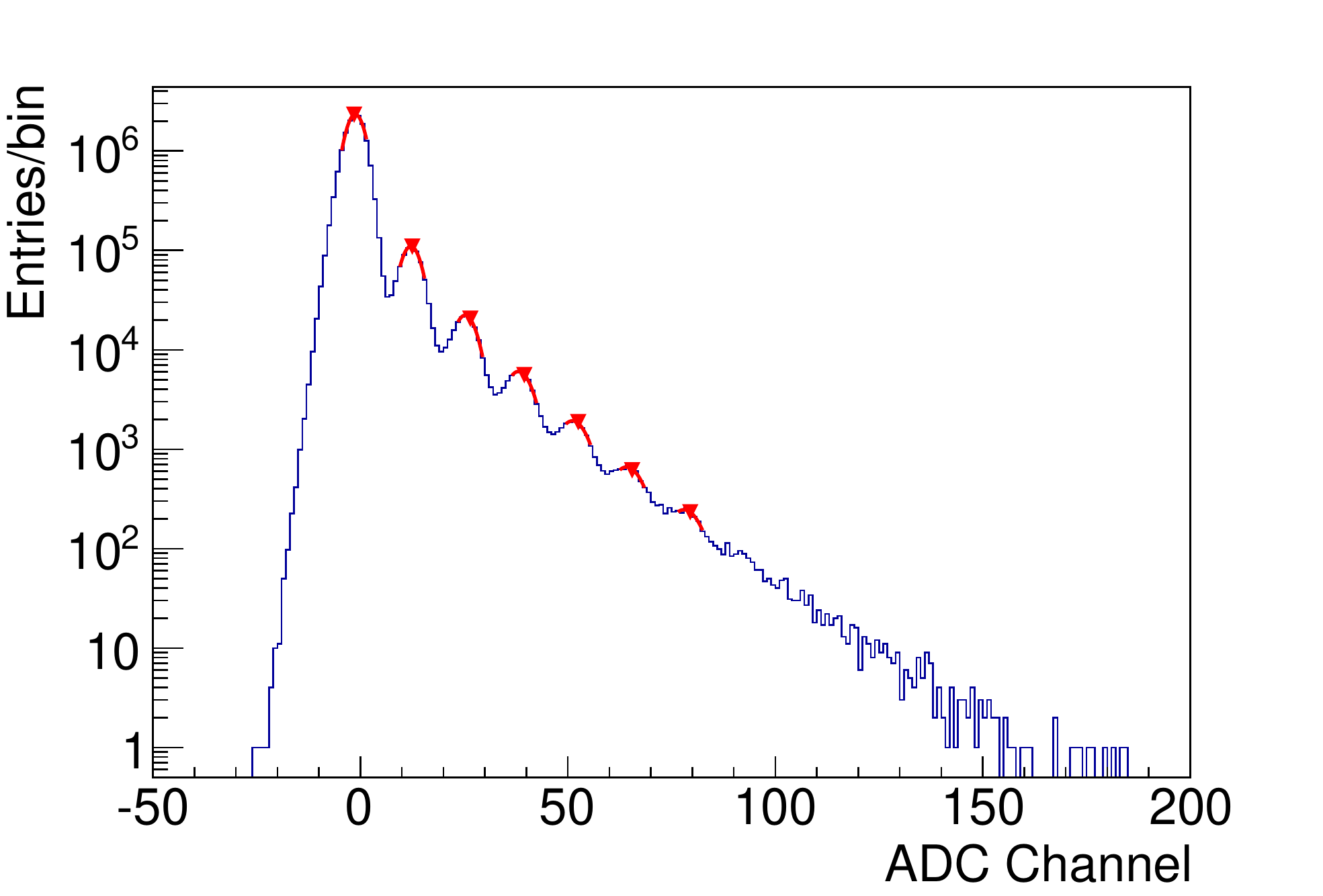}
    \includegraphics[width=0.49\textwidth]{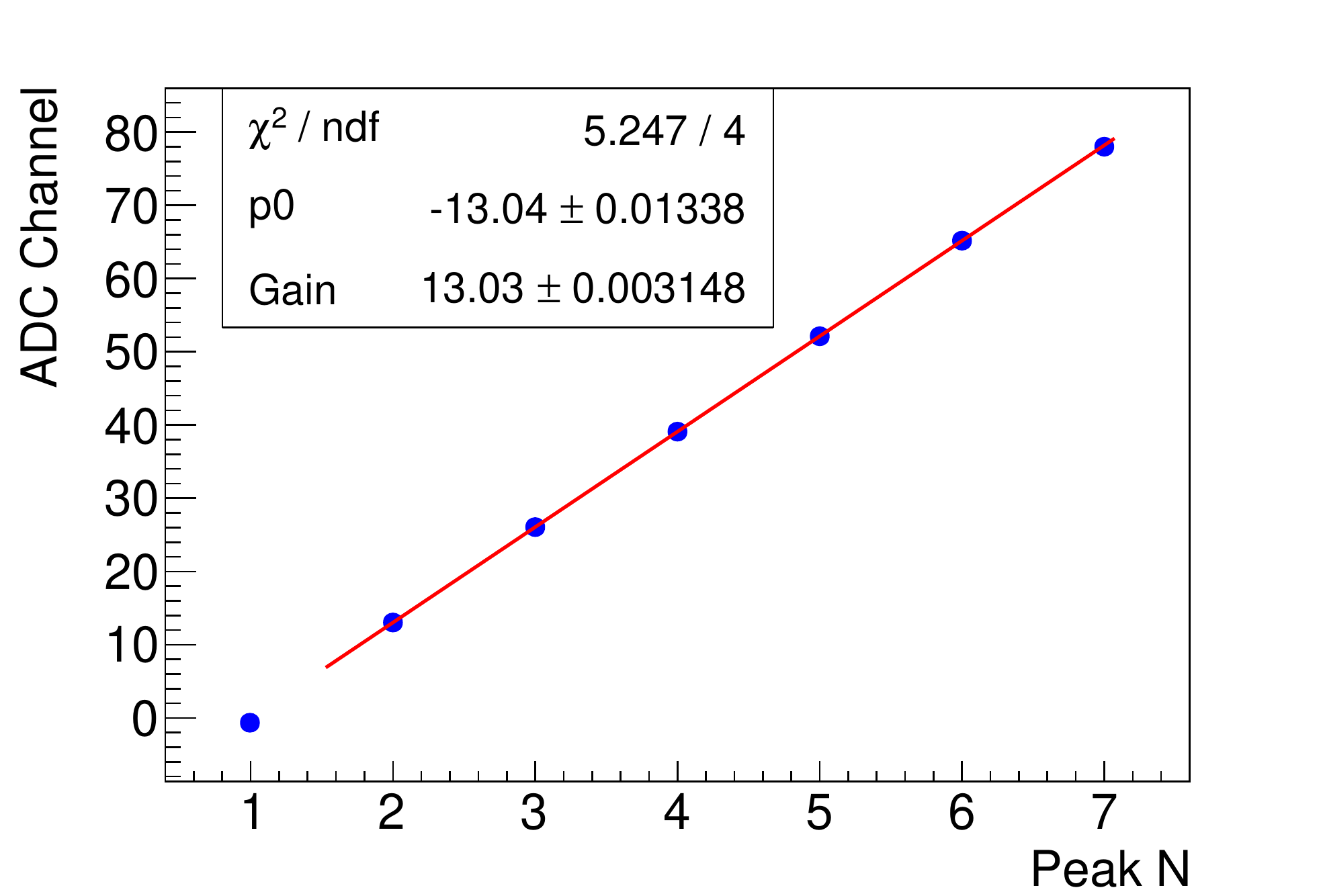}
    \caption{Left: Single Photon Spectrum (SPS) of a SiPM obtained with dark current events. Individual photoelectron peaks are identified and fitted to a Gaussian. Right: Average number of ADC counts produced by different number of photoelectrons and linear fit obtaining the conversion gain. The point corresponding to the pedestal is excluded from the fit.}
    \label{fig:DK}
  \end{center}
\end{figure}

Before data taking, the conversion gain of each individual sensor must be determined. The PMTs are calibrated using their single photon response using a 400~nm wavelength LED mounted in the centre of the tracking plane, as described in \cite{Alvarez:2013gxa}. The SiPMs are calibrated using the dark current of the sensors. A number of pixels within each channel are fired within the read-out integration time of 1~$\mu$s due to thermally generated photoelectrons. Recording the signal level per $\mu$s results in a spectrum of the form shown in figure \ref{fig:DK}-\emph{left}. By fitting a Gaussian to the peaks corresponding to different numbers of photoelectrons and using the centroid position in a linear fit allows for a determination of the conversion gain (an example is shown in figure \ref{fig:DK}-\emph{right}).

\subsection{Data} 
\label{subsec:DataSets}
\begin{figure}
  \begin{center}
    \includegraphics[width=0.7\textwidth]{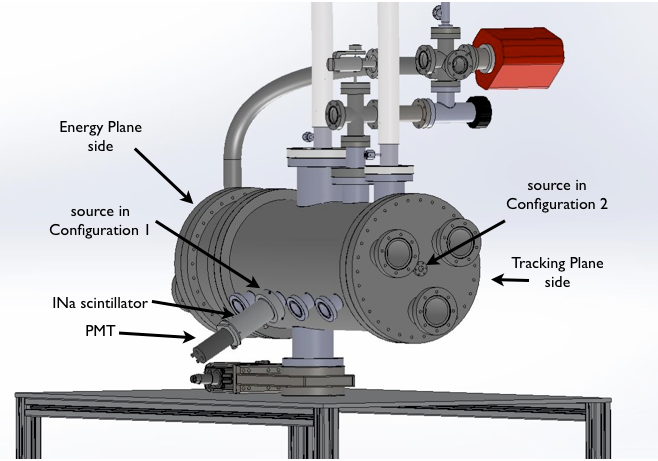}
    \caption{Schematic of NEXT-DEMO showing the different source positions for the data taking.}
    \label{fig:source}
  \end{center}
\end{figure}

The analyses presented in this paper used two separate datasets. Both proceed from the interaction of 511~keV gammas produced by the annihilation of positrons emitted by a \NA\ source in the detector volume but differ in the source position (figure \ref{fig:source}). In \textsl{Configuration 1} the \NA\ source was located between one of the transparent lateral ports of the vessel and a NaI scintillator placed outside. In this configuration, read-out was triggered by coincidence between a S1 and a pulse in the external NaI scintillator. This trigger is possible because of the back-to-back photons produced in positron annihilation.  In \textsl{Configuration 2}, the \NA\ source was located centered in one of the end caps of the vessel, axis Z = 0. The trigger required the coincidence of an S1-like signal in at least 3 central PMTs.

\subsection{Monte Carlo data generation} 
\label{subsec:MC}
A set of Monte Carlo end-to-end simulated events of the same kind of data has been used throughout this analysis, both as a tool for reconstruction and as a check of the appropriateness of the methods. A detailed simulation of the NEXT-DEMO prototype has been developed in NEXUS, the Geant4-based \cite{Agostinelli:2002hh,Allison:2006ve} simulation software of the NEXT collaboration. 
This Monte Carlo dataset was generated simulating the two 511~keV back-to-back gammas coming from the annihilation of the positron emitted in the decay of a \NA\ nucleus and the 1274~keV gamma emitted in the de-excitation of the daughter nucleus. Such particles are generated in the place where the radioactive source is located in NEXT-DEMO and are propagated through the materials of the detector, where all the relevant processes are taken into account. The ionization electrons resulting from the ionization of the gas are drifted through the active region and, when they enter the electroluminescent region, a secondary scintillation signal is simulated, according to previously produced lookup tables, which give the response of the sensors (both PMTs and SiPMs) to the electroluminescent light generated in a particular point of the EL region. The response of the sensors, in photoelectrons, is digitized, adding fluctuations in the gain, electronic noise, and shaping according to measurements taken in NEXT-DEMO. 

\subsection{Data preprocessing and selection} 
\label{subsec:Data}
\begin{figure}
  \begin{center}
    \includegraphics[width=0.49\textwidth]{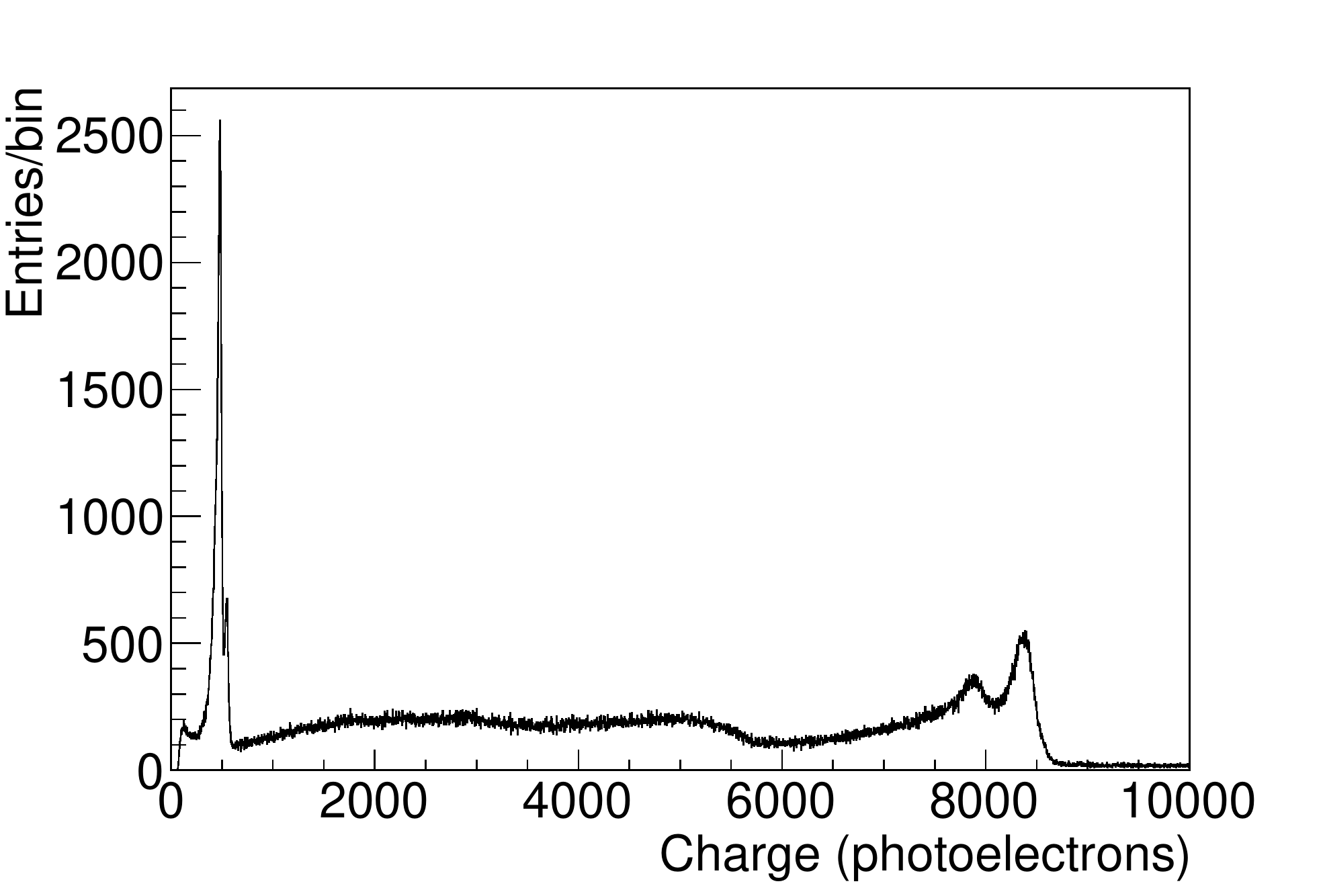}
    \includegraphics[width=0.49\textwidth]{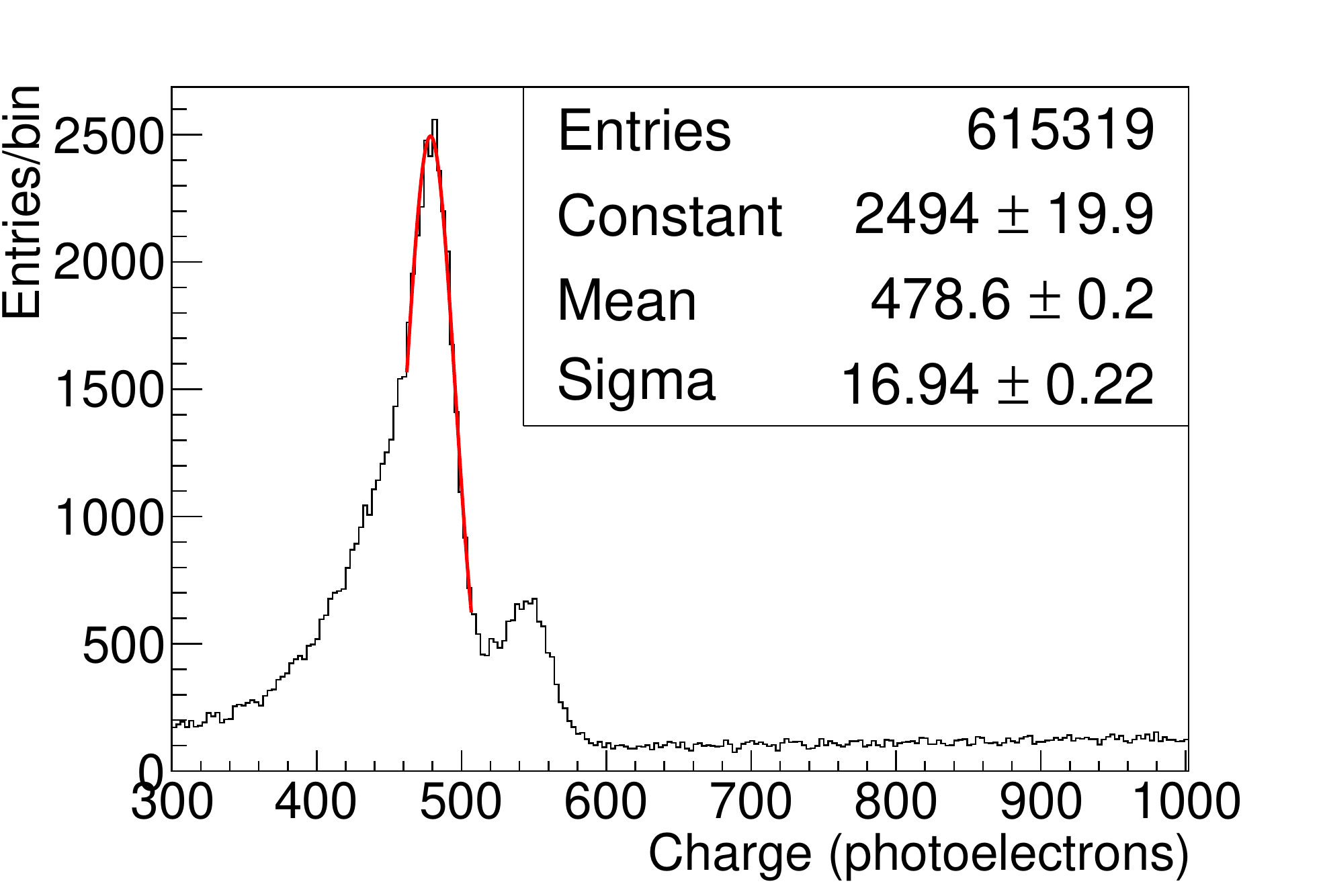}
    \caption{Left: Energy spectrum of  \NA\ source before any correction. Right: Gaussian fit to X-ray peak for X-ray event selection.}
    \label{fig:CathodeSum}
  \end{center}
\end{figure}

Raw data first passes through a data preprocessing algorithm common to all analyses. This applies pedestal correction to all channels before identifying S1-like and S2-like signals and rejecting any events with multiple S1-like peaks or where the S2 signal is less than 20 photoelectrons. Figure \ref{fig:CathodeSum}-\emph{left}, shows the signal induced by \NA\ averaged over all 19 PMTs. This is the typical \NA\ energy spectrum, where the photoelectric and escape peaks as well as the Compton edge are clearly visible along with the Xe X-ray peaks. Events with energy within 1 sigma of the most prominent X-ray peak ($K_\alpha$) are considered to be due to the interaction of these X-rays and constitute the basic dataset for the subsequent analyses (figure \ref{fig:CathodeSum}-\emph{right}). The purity of X-ray events in this initial selection is high in such a way that less than 5\% within this range are not X-ray events, corresponding to the low energy distribution of the Compton effect (flat region in figure \ref{fig:CathodeSum}-\emph{right}).

\subsection{Position reconstruction} 
\label{subsec:PosRec}
\begin{figure}
  \begin{center}
    \includegraphics[width=0.9\textwidth]{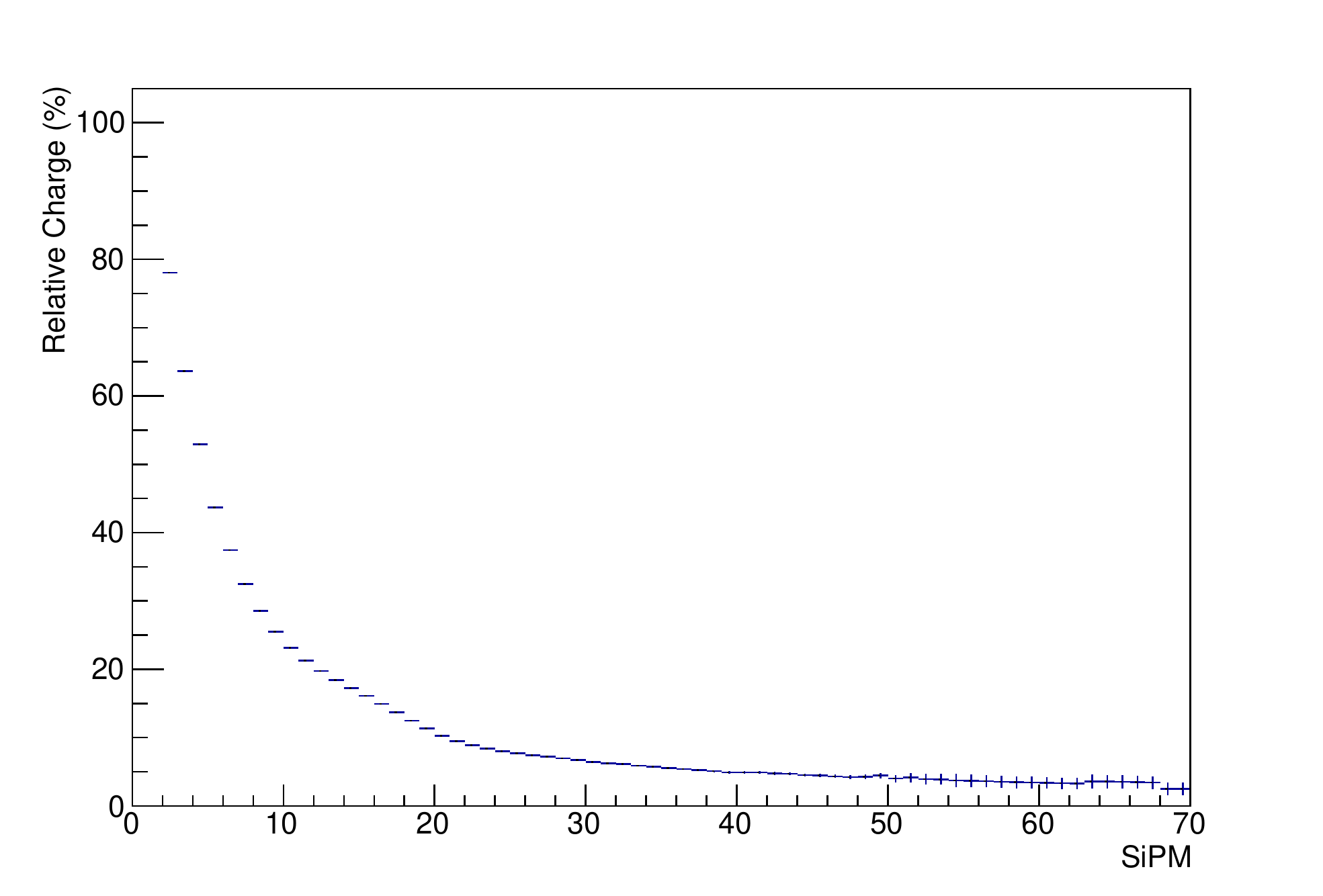}
    \caption{Relative charge to maximum signal in the tracking plane. The plot shows how relative charge decreases until an asymptotic value of 10\% of the maximum charge.}
    \label{fig:HistRel}
  \end{center}
\end{figure}

Spatial reconstruction of the deposited energy in the detector is performed using the S2 signal produced in the EL region using the information available from the tracking plane. The EL production is isotropic but for losses in the grids and, as such, a reconstruction of the position is made using the barycentre of the deposited charge. A preselection of the useful channels is made considering the relative charge compared to that of the maximum channel. As can be seen in figure~\ref{fig:HistRel}, the charge observed in each channel tends to decrease asymptotically to a level of 10\%. Taking this as the noise baseline, the barycentre is calculated using only those channels with charge  greater than 10\% of that of the channel with greatest charge.

\begin{figure}
  \begin{center}
    \includegraphics[width=0.7\textwidth]{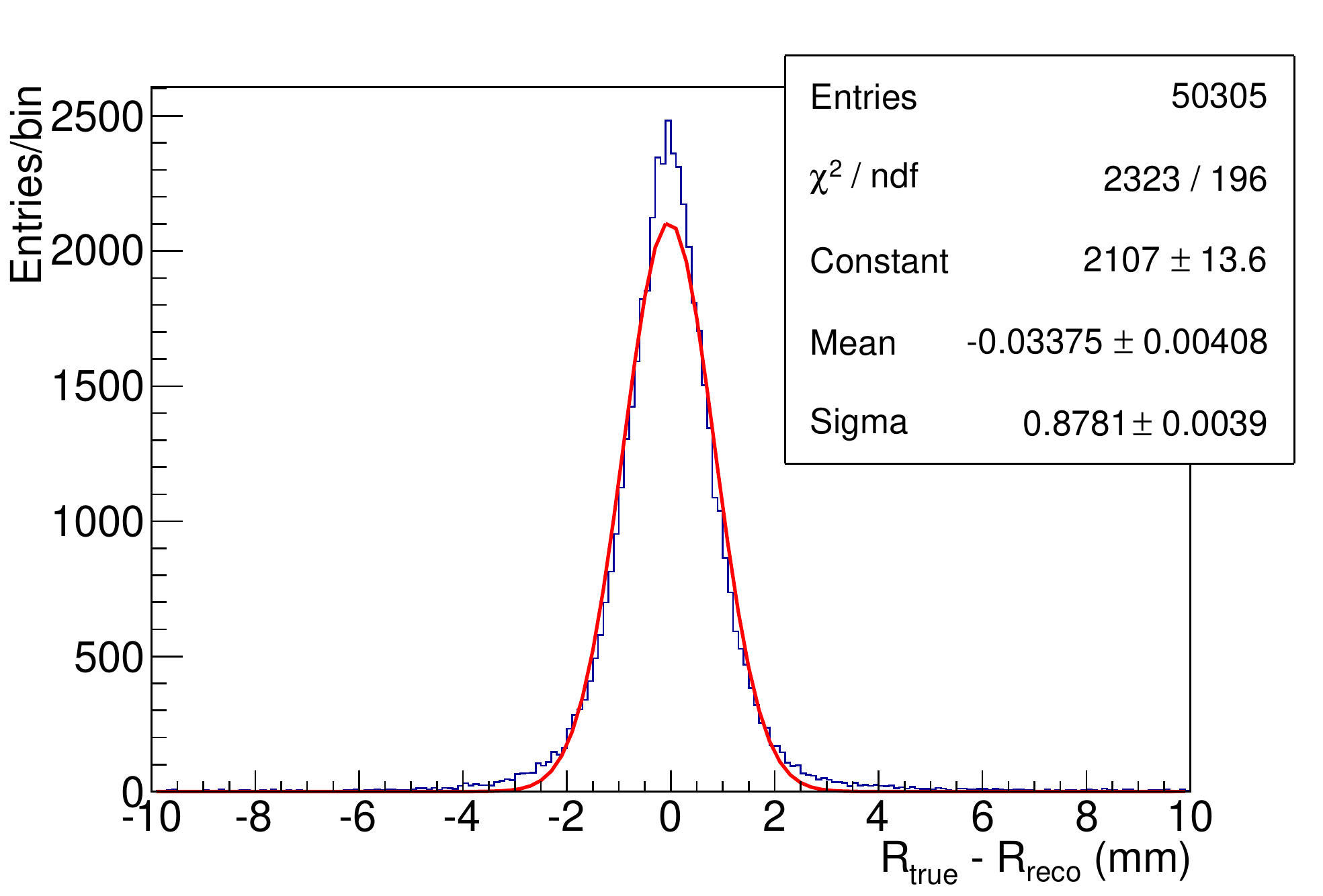}
    \caption{Distribution of the reconstructed position. $R_{true}$ is the true position of events given by MC and $R_{reco}$ is the calculated position.}
    \label{fig:Pull}
  \end{center}
\end{figure}

This assumption has been checked using NEXUS under the same conditions. The barycentre is calculated as described above using the SiPM channels with charge of greater than 10\% of the SiPM with maximum charge and then compared to the averaged position of the energy deposits recorded by Geant4. Figure \ref{fig:Pull} shows the distribution on the reconstructed position where $R_{true}$ is the true position of events given by MC and $R_{reco}$ is the calculated position. In the data the statistics reduce significantly at $R_{reco} > 60$~mm due to set trigger conditions favor the events in the center of the detector volume, for this reason the fiducial region is defined by this value. As can be seen in figure \ref{fig:Pull}, the position is reconstructed accurately within this region. The z fiducial region depends on the particular analysis, with those analyzing drift effects requiring events produced closer to the grids than are useful for the resolution analysis of section \ref{subsec:Extrapol}.

\section{Detector response using x-rays} 
\label{sec:DetRes}
X-ray events are particularly useful to monitor the performance of the detector and to calibrate for effects beyond the scope of the basic conversion gain because of their point-like nature and their abundance in source based data taking. This section describes the development of methods with an eye on their extrapolation to the larger detectors which will make the physics measurements of~NEXT.

\subsection{Tracking Plane Response} 
\label{subsec:TPR}
As explained in \cite{Alvarez:2013gxa}, the \textsl{tracking plane} consists of 256 Hamamatsu S10362-11-050P SiPMs distributed between 4 boards, each with 64 sensors with common bias voltage, and is positioned 2 mm behind the EL production region. Prior to its introduction in the TPC, a dedicated study of the sensors was done, with a final spread in gain below 4\% at room temperature \cite{Alvarez:2012bbb}. However, due to the posterior addition of a wavelength shifter coating (TPB) over the SiPMs \cite{Alvarez:2012ub}, to make them sensitive to the xenon scintillation light, together with the addition of the electronic read out chain, a regular monitoring of the SiPM response is needed to maintain the spread in gain constant.

\subsubsection{X-ray calibration monitoring} 
\label{subsubsec:XRhomog}
The calibration constants of the individual SiPMs are determined as described in section \ref{subsec:CalSen}. However, during long runs there is the possibility that channels could become faulty or that the TPB on the surface of the SiPMs could degrade, causing a change in the conditions of the detector. The relative calibration of the sensors can be monitored using the xenon $K_\alpha$ deposits.

\begin{figure}
  \begin{center}
    \includegraphics[width=0.49\textwidth]{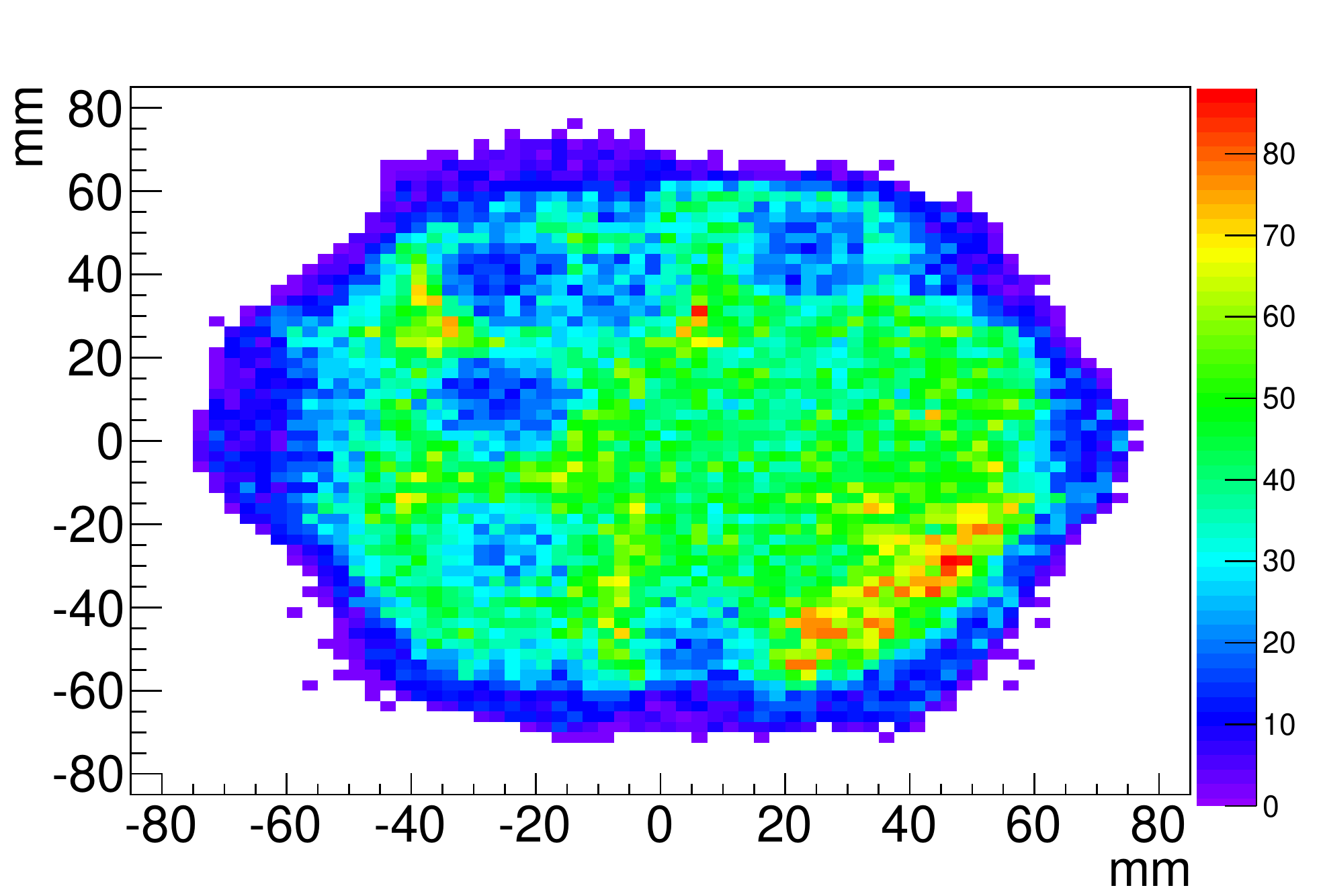}
    \includegraphics[width=0.49\textwidth]{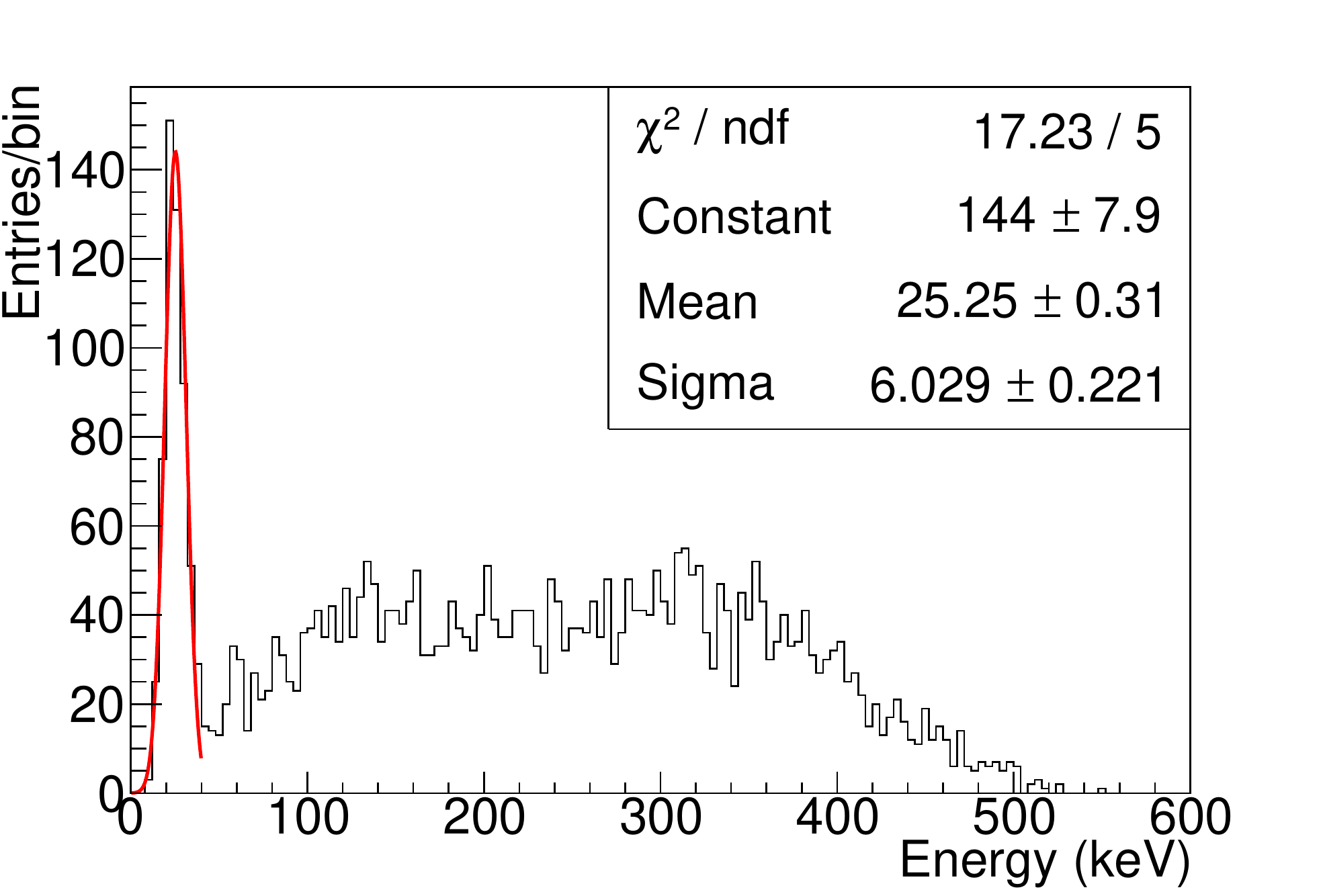}
    \caption{Left: 2D projection of the spatial distribution of x-rays inside the detector. Right: "Low energy" spectrum of \NA\ obtained with one SiPM of the \textsl{tracking plane}, and gaussian fit to the X-ray peak. }
    \label{fig:XR}
  \end{center}
\end{figure}

Under the trigger conditions of the data considered here around 15\% of the events are caused by xenon X-ray interactions. As can be seen in figure \ref{fig:XR}-\emph{left}, the events are distributed across the full volume of the detector in x-y. Due to the smallness of these events and the distance where SiPMs are located, a few millimetres away from EL light generation, the light produced tends to be concentrated in few channels. Considering as estimator of the energy released in each recorded event only the channel with most charge, a low energy spectrum is achieved (example in figure \ref{fig:XR}-\emph{right}) which exhibits a peak in the region of the X-ray energy. The relative position of this peak can be used to effectively monitor the SiPM calibration and any change over data taking taken into account in analysis.

\subsection{Energy Plane Response} 
\label{subsec:EPR}
The measurement of event energy is affected by a number of processes in the detector, many of which can be corrected with appropriate calibration. The individual gains are well determined by the LED calibration but there remain inhomogeneities in the averaged response due to differences in PMT quantum efficiency, and to the imperfect rotational symmetry of the light tube and variation in the amount of deposited TPB across the detector. The first of these effects is limited by preselecting PMTs with little variation in response and can be taken into account in the calculation of the averaged signal as described in section \ref{subsubsec:QE}. The latter effects cannot be separated but their combined effect can be calibrated for using the methods described in section \ref{subsubsec:XYCorr}.

\subsubsection{Determination of relative response of the PMTs} 
\label{subsubsec:QE}

Disparity in the response of the photosensors that make the energy measurement (PMTs) can have an impact in the energy resolution of the detector \cite{Nygren:2009NIM}. The relative response of the PMTs can be measured using the x-ray signal. Considering the response function of a PMT as being described by:

\begin{equation}
  q_{i(K_\alpha)} = N_{\gamma}\cdot{}TF_{i}(x,y)\cdot{}QE_{i}
\end{equation}

where $q_{i(K_\alpha)}$ is the recorded charge in photoelectrons of the PMT with index $i$ due to a $K_\alpha$ deposit, $N_{\gamma}$ is the number of photons produced by de $K_\alpha$ deposition, $QE_{i}$ represents the quantum efficiency, and $TF_{i}(x,y)$ is the transmission function of the light tube, which is a map which describes the probability that a photon generated at position (x,y), will pass through the photocathode of the PMT $i$. 

Using the Monte Carlo data described in section \ref{subsec:MC} and setting the gains and $QE$ of all PMTs to 1, $TF_{i}(x,y)$ can be obtained as the only effect which contributes to differences in the PMTs' response.
\begin{equation}
  q_{i MC(K_\alpha)} = N_{\gamma}\cdot{}TF_{i}(x,y)
  \label{eq:QEMC}
\end{equation}
A determination of $TF_{i}(x,y)$ coupled with a good knowledge of the individual conversion gains of the PMTS, the relative $QE$ between the different PMTs can be established. Taking the response of the central PMT as reference it can be seen that the relative $QE$ can be determined as:
\begin{equation}
  \frac{QE_{i}}{QE^{*}} = \frac{q_{i(K_\alpha)}TF^{*}(x,y)}{q^{*}_{(K_\alpha)}TF_{i}(x,y)}
  \label{eq:QErel}
\end{equation}
where $q^{*}_{(K_\alpha)}$ is the response in photoelectrons of the central PMT and $QE^{*}$ its quantum efficiency.

\begin{figure}
  \begin{center}
    \includegraphics[width=0.49\textwidth]{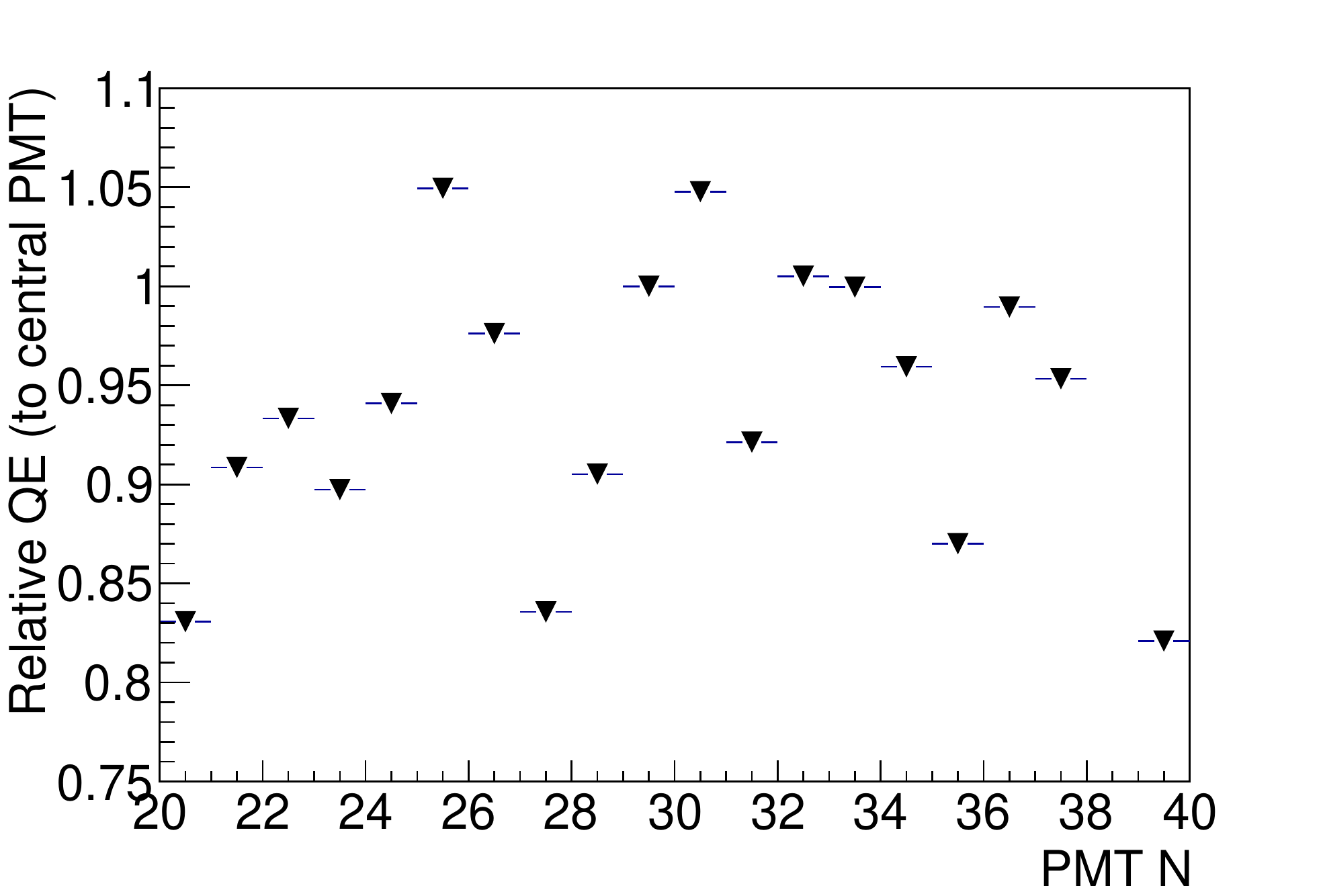} 
    \includegraphics[width=0.49\textwidth]{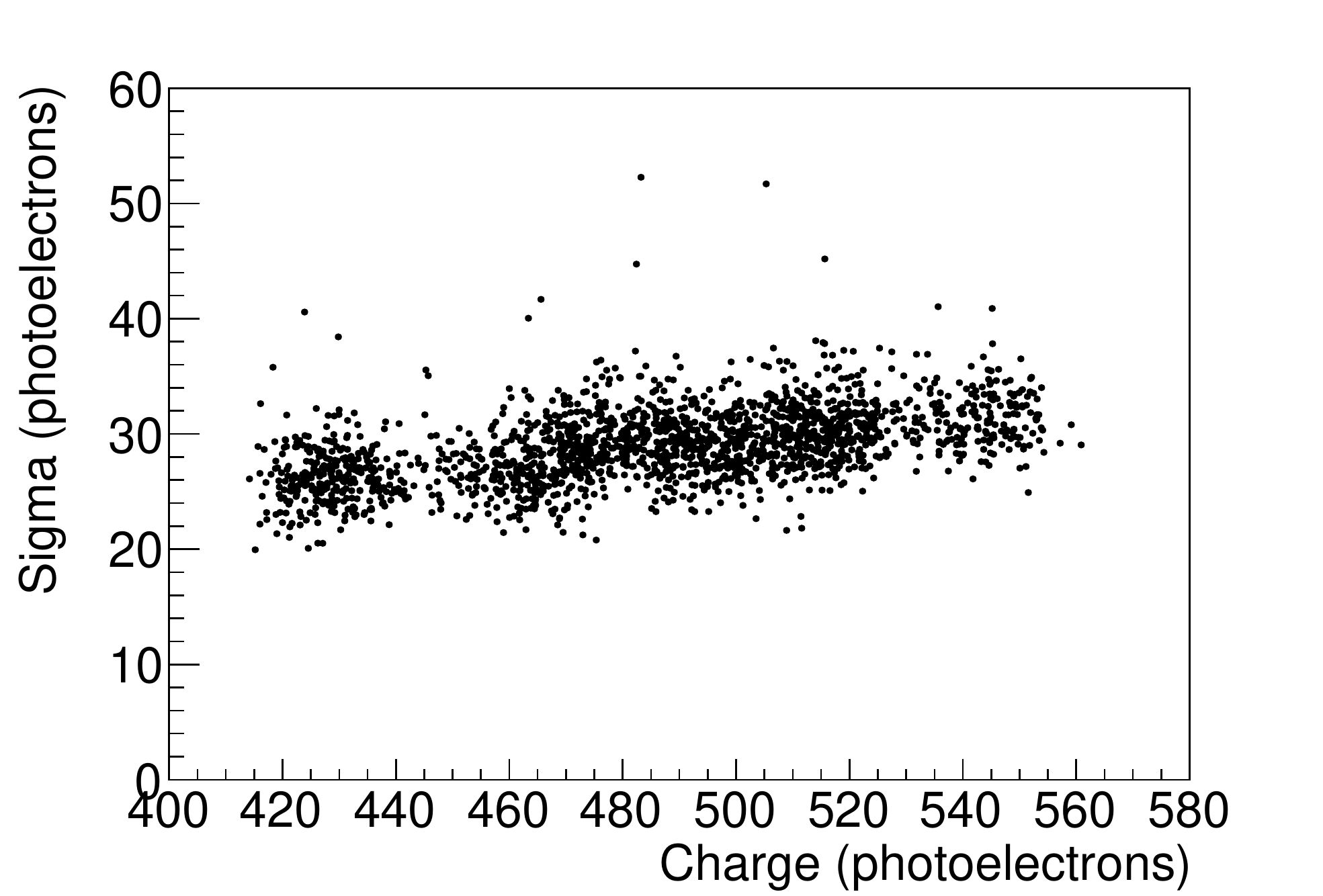}
    \caption{Left: Relative quantum efficiencies of the 19 PMTs of the TPC, referenced to the central one. Right: The fitted sigma of the 19 PMTs for all XY bins. Isolated points at upper part correspond to bins with lower statistics.}
    \label{fig:QErel}
  \end{center}
\end{figure}

The response of the detector to the $K_\alpha$ X-rays was analysed in bins of 1~cm$^2$ and the resulting distributions for each PMT, in each bin, were fitted to a Gaussian distribution and the mean and sigma extracted. The mean values were then used to calculate the relative $QE$. The spread can be seen in figure \ref{fig:QErel}-\emph{left} where all but 4 of the PMTs exhibit $QE$ within 10\% with the outliers within 20\%. This level of agreement allows for a consideration of a weighted sum as an estimator of the energy deposited in the detector. In addition, the sigma values (figure \ref{fig:QErel}-\emph{right}) were used to calculate the variance of each distribution and its inverse were taken as the weight of a PMT within that $(x,y)$ bin and used in the calculation of the weighted sum (as described in section \ref{sec:EnergyResolution}). The means are used to calculate geometric correction factors as described in section \ref{subsubsec:XYCorr}.

\subsubsection{Quantum Efficiency degradation} 
\label{subsubsec:QEdeg}

Continuous exposure to VUV light can damage the photocathode of a PMT. In NEXT-100, PMTs will be located behind TPB coated sapphire windows, effectively blocking the PMTs from VUV. Simulations of this geometry using a silicone gel to optically couple the the sapphire and PMT windows show no significant reduction in detected light and that the X-ray emissions are still easily discernible. However, in the NEXT-DEMO detector, the direct exposure to the xenon scintillation light can cause a degradation of the photocathode detection efficiency.

\begin{figure}
  \begin{center}
    \includegraphics[width=0.7\textwidth]{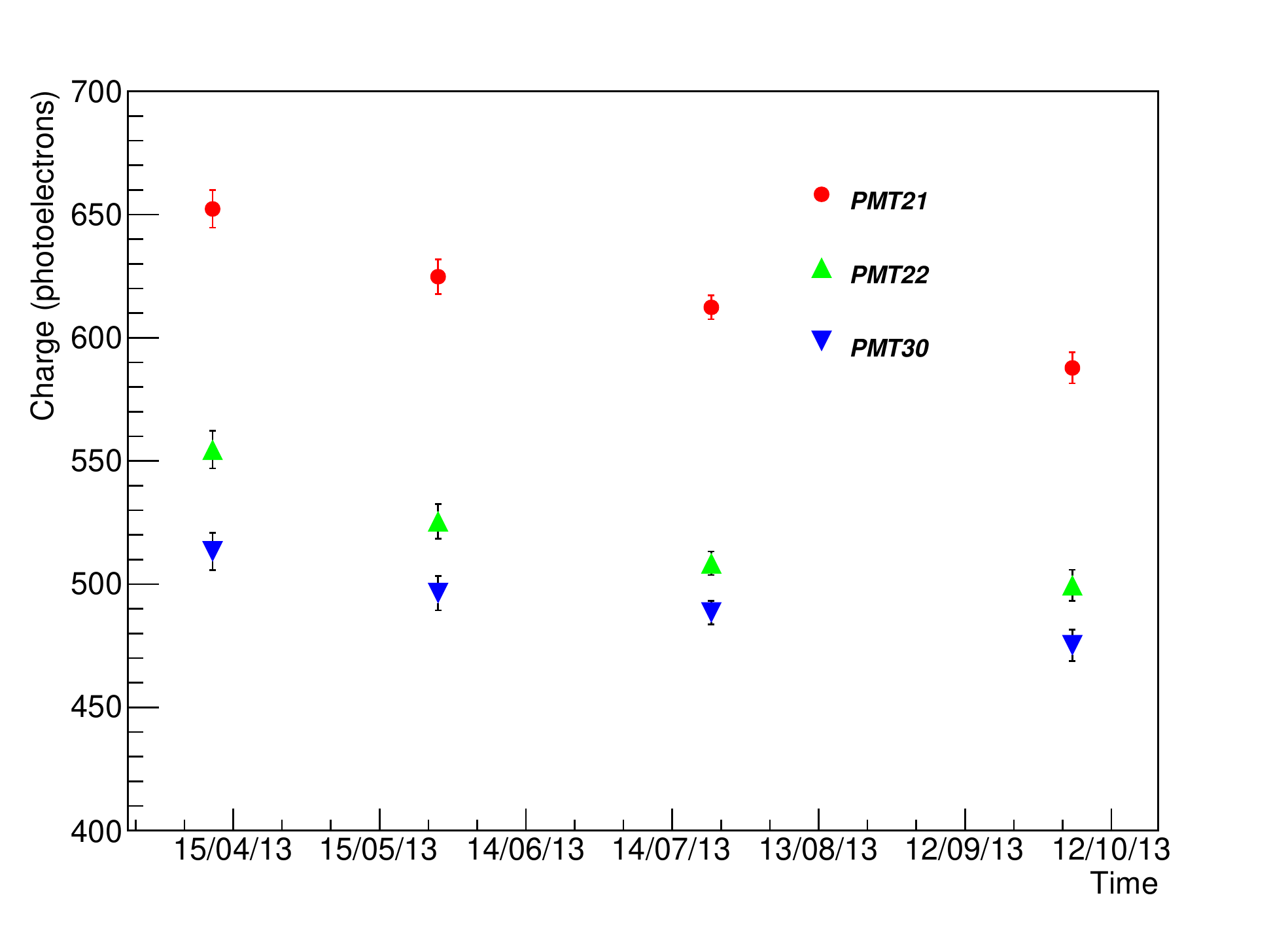}
    \caption{Average response of 3 of the 19 PMTs to $K_\alpha$ X-rays as a function of time. A reduction in response is observed.}
    \label{fig:QEdeg}
  \end{center}
\end{figure}

This effect can be observed as a time dependent reduction of the detected $K_\alpha$ X-ray charge. Figure \ref{fig:QEdeg} shows the mean position in charge of the $K_\alpha$ peak over a period of 6 months. As can be seen, there is a degradation of the charge with time.

This degradation can be modelled as described in equation \ref{eq:QEdeg}:
\begin{equation}
  q_{i(K_\alpha)} = N_{\gamma}(K_\alpha) \cdot{} PDE_{i}(\alpha t)
  \label{eq:QEdeg}
\end{equation}
where $N_{\gamma}(K_\alpha)$ is the number of photons produced by the $K_\alpha$ deposition, and $PDE_{i}(\alpha t)$ is the photon detection efficiency of the PMT, parameterized as linearly dependent on time with gradient $\alpha$. $\alpha$ was determined to be a fraction lost of charge of $\sim$ 0.06$\%$ per day.

\subsubsection{XY Response} 
\label{subsubsec:XYCorr}
\begin{figure}
  \begin{center}
    \includegraphics[width=0.49\textwidth]{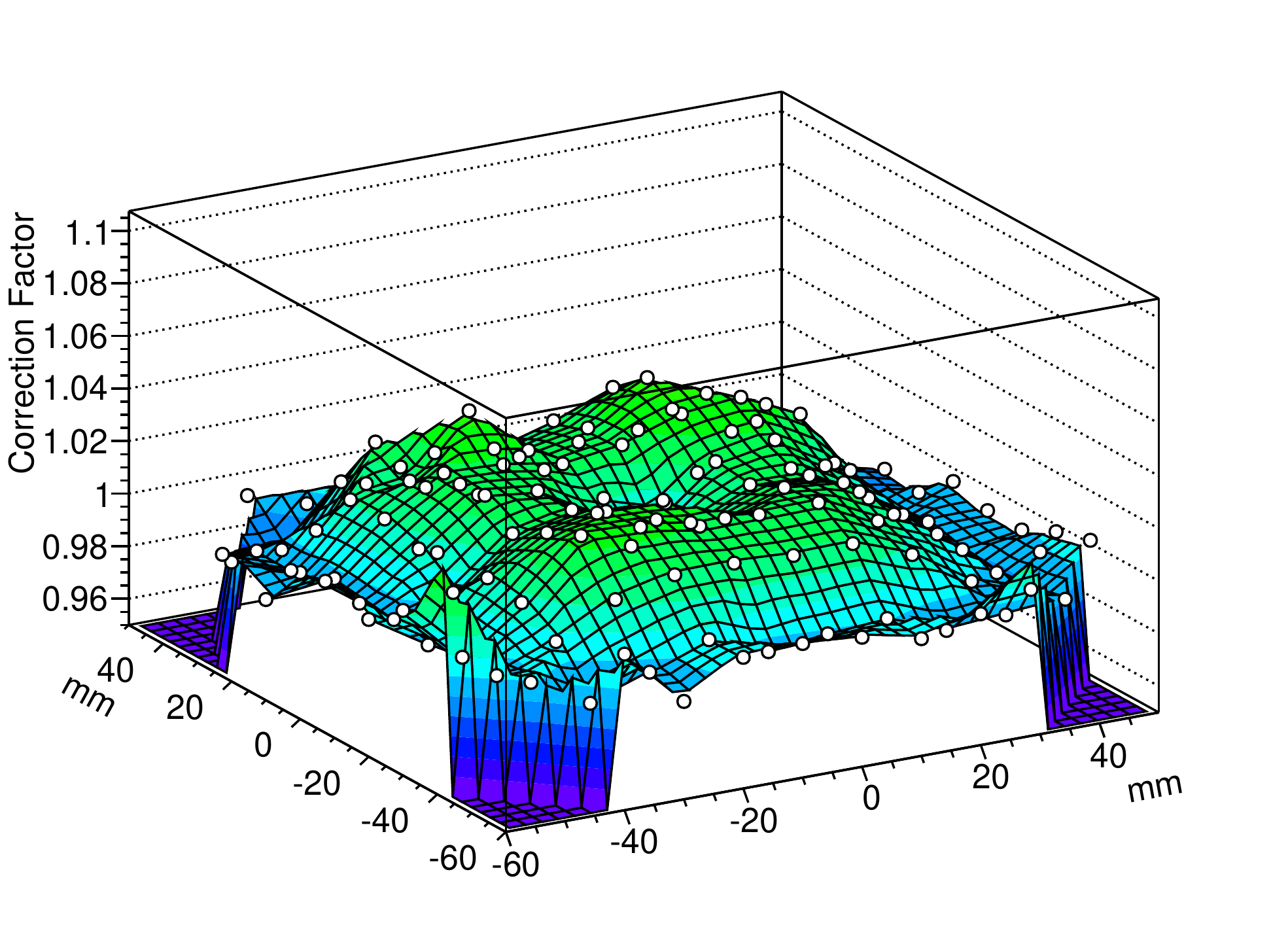} 
    \includegraphics[width=0.49\textwidth]{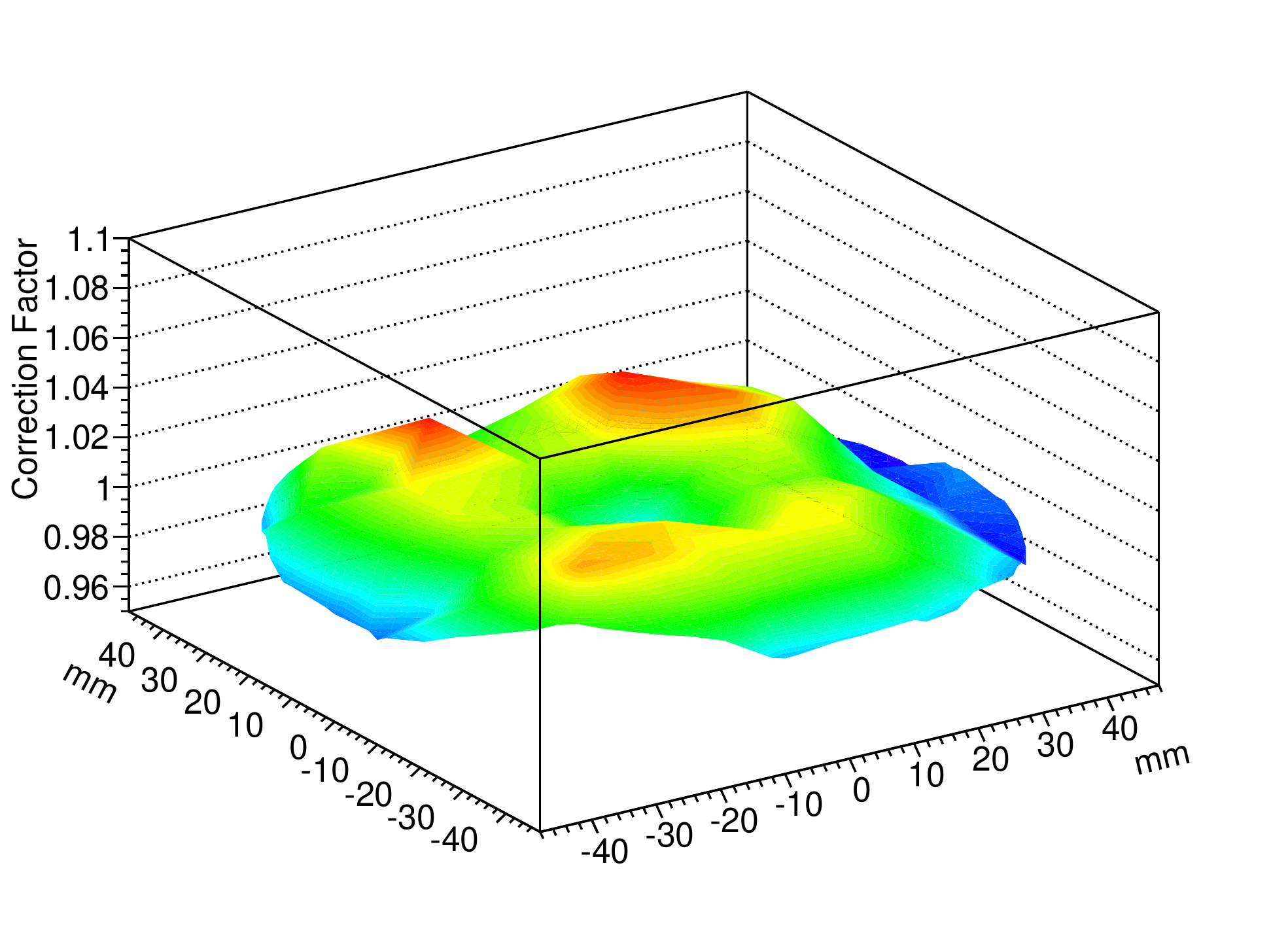} 
    \caption{Left: Original correction factors (white dots) with Delaunay interpolation map. Right: Delaunay triangulation of the grid increasing the number of correction factors by a factor $10^{2}$.}
    \label{fig:DelInt}
  \end{center}
\end{figure}

After taking into the account the variation in response of the PMTs in the weighted sum, there still remains a dependence on event position in the detector response. Using the mean position of the $K_\alpha$ deposits as calculated in the previous section, a correction factor can be determined by normalizing to the central bin (figure \ref{fig:DelInt}-\emph{left}). The resolution of the grid can then be improved using a Delauney triangulation~\cite{Delaunay} between the bins, increasing the number of factors by 2 orders of magnitude (figure \ref{fig:DelInt}-\emph{right}).
These correction factors can then be introduced as an additional term in the weighted sum, as described in section \ref{subsec:Extrapol}.

\section{Properties of Xenon EL-based TPC} 
\label{sec:PropXTPC}
As explained in section \ref{sec:Introduction}, EL TPCs have huge potential in the field of \bbonu\ physics. Good energy resolution can be achieved and topological reconstruction can be used to reduce backgrounds. However, achieving the optimum performance requires a deep understanding of the detector response so that inhomogeneities in energy reconstruction and blurring of the event topology do not reduce sensitivity. The point-like nature of X-ray induced events means that they can be used to monitor fundamental properties of the gas and detector.

Of particular interest are the properties of the EL gap. Since electrons continuously produce light as they pass between the gate and anode, a point-like deposition will be smeared out in three dimensions at the read-out plane. These effects are convoluted with the diffusion of the charge cloud during drift which can also be studied, along with the drift velocity, using the $K_\alpha$ X-ray deposits.

\subsection{Drift Velocity} 
\label{subsec:DriftVel}
\begin{figure}
  \begin{center}
    \includegraphics[width=0.49\textwidth]{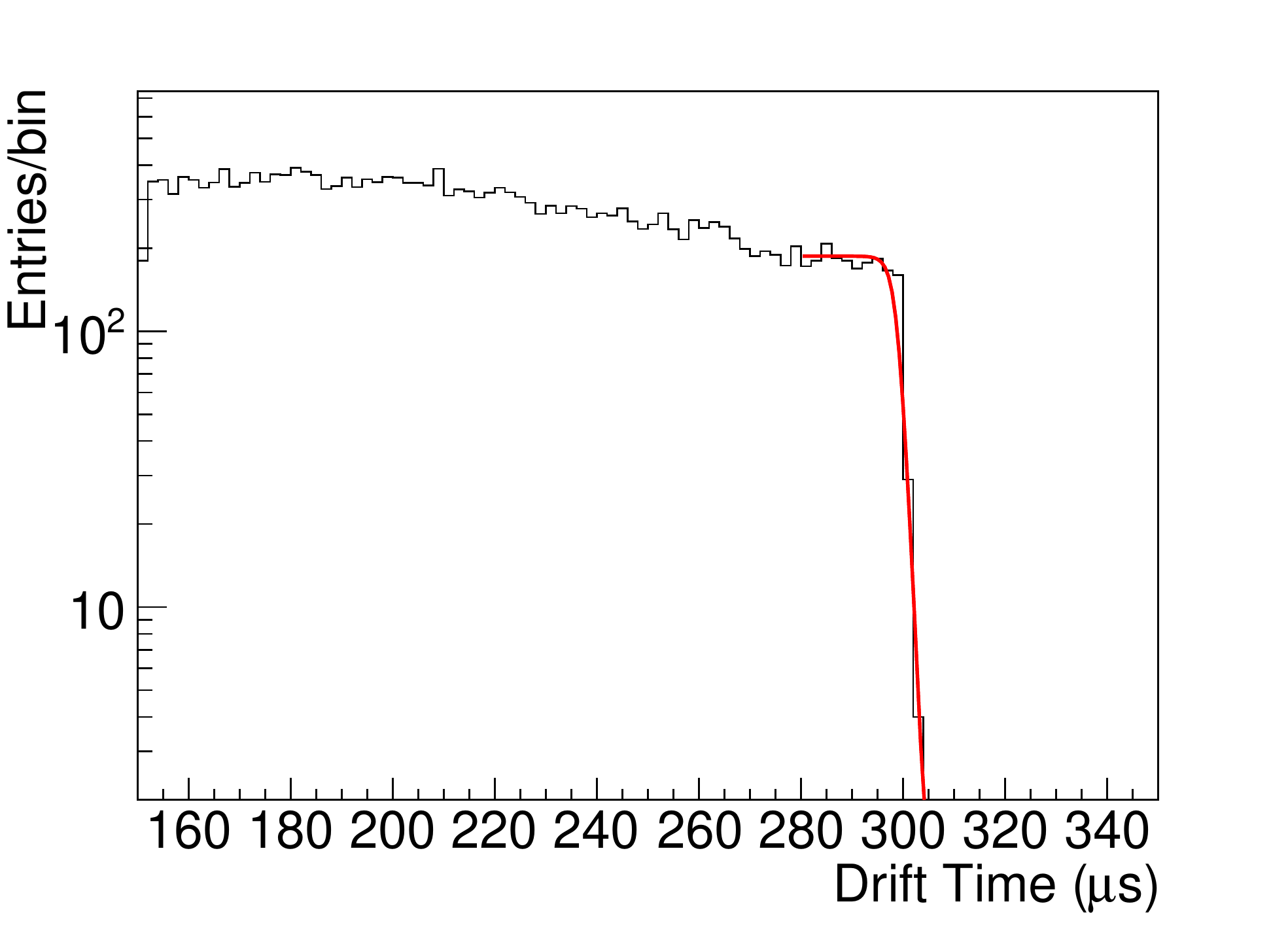} 
     \includegraphics[width=0.49\textwidth]{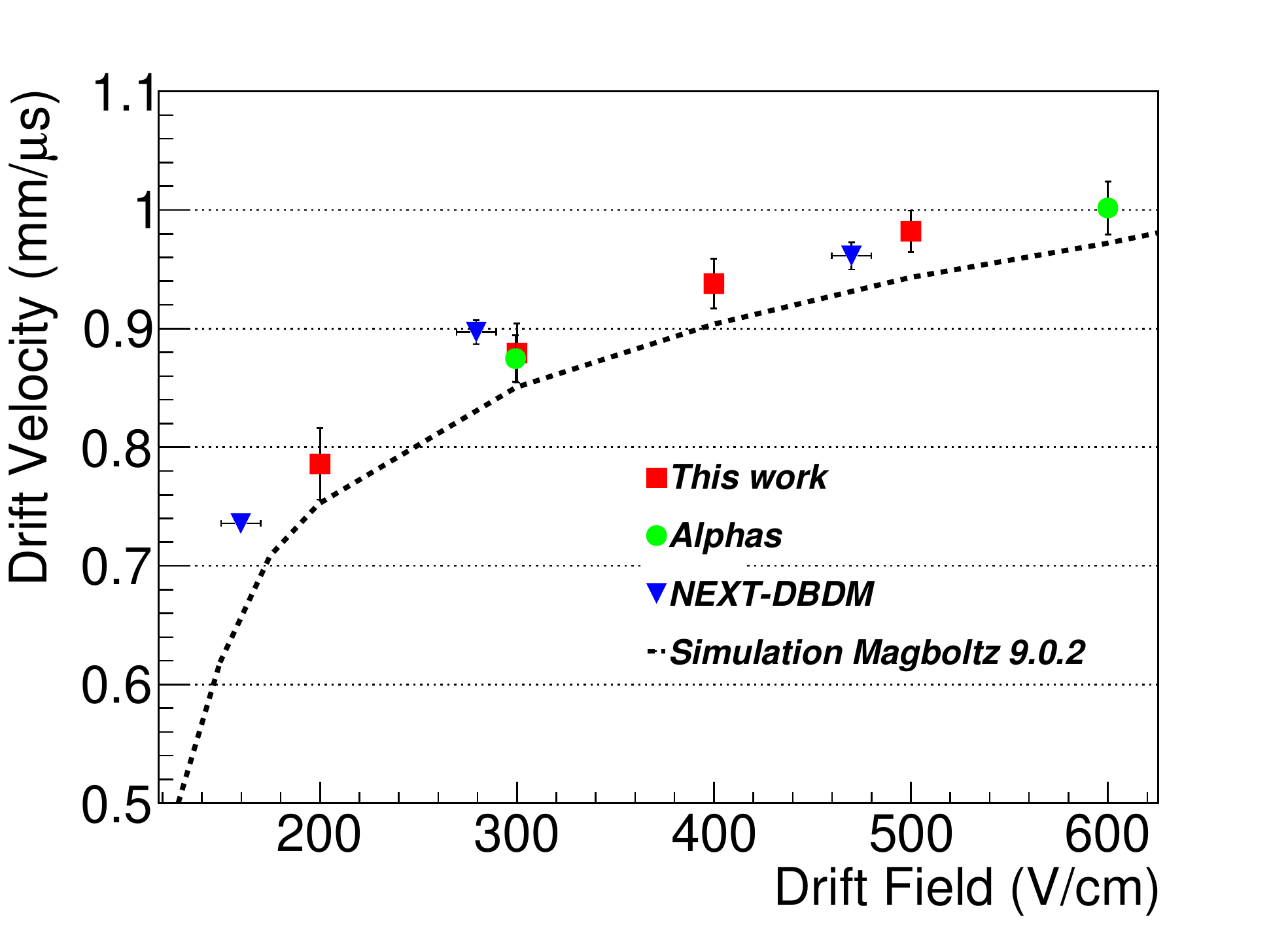} 	
    \caption{Left: Event time distribution for 0.5~kV~cm$^{-1}$ drift field, and its fit to the Heaviside function. Right: Drift velocity as a function of drift field, for xenon gas at 10~bar. The red square points are the measured values, green circles correspond to the results of \cite{Alvarez:2012hu}, blue triangles are from \cite{Alvarez:2012kua} while the dashed curve is the prediction for pure xenon at 10~bar from the Magboltz 9.0.2 simulation.}
    \label{fig:DriftVelocity}
  \end{center}
\end{figure}

Electron drift velocity ($v_{d}$) can be determined analyzing the longitudinal event time distribution in the TPC. The drift time ($t_d$) is well defined by the difference in detection time between the S1 and S2 signals in the events selected as X-ray induced. There exists a maximum drift time ($t_{dmax}$) corresponding to the events just inside the drift region next to the cathode. This maximum can be determined as the half-maximum of a Heaviside function fitted to the event time distribution (shown in figure \ref{fig:DriftVelocity}-\emph{left}). The maximum drift distance can be calculated from the detector design parameters as the total drift distance plus half of the width of the EL region, since the peak of light production is well estimated by that point. In NEXT-DEMO, these values are 300~mm and 2.5~mm respectively~\cite{Alvarez:2012xda}. 

Drift velocity is lower in drift region than in EL region, however, due to smallness of the last, we assume this difference negligible compared to the total drift distance. The drift velocity can be then calculated from the ratio between the maximum drift distance $D_{dmax}$ and the maximum drift time $t_{dmax}$: 
\begin{eqnarray}
  D_{dmax} =& 300 + (5/2) &= 302.5~\mbox{mm}\\
  v_{d} =& \displaystyle\frac{D_{dmax}}{t_{dmax}} &= \frac{302.5~\mbox{mm}}{t_{dmax}}
  \label{eq:vd}
\end{eqnarray}

\textsl{Configuration 1} data were used to determine the drift velocity in order to maximise the number of events near the cathode. The drift velocity was determined for 4 drift field settings (0.5, 0.4, 0.3, 0.2 kV~cm$^{-1}$). 

The results are shown in figure \ref{fig:DriftVelocity}-\emph{right}, where they are compared to the expectation using electron scattering cross-sections for xenon at 10~bar obtained with the simulation based on version 9.0.2 of Magboltz. The difference between data and simulation may be due to the presence in the gas of xenon clusters, although not being stable, reduce the gas mix density increasing the electron drift velocity as well as the uncertainties in the cross sections used by the simulated model. These results are in agreement  with previous data obtained by the \textsl{NEXT Collaboration} as published in~\cite{Alvarez:2012kua, Alvarez:2012hu}.

\subsection{Longitudinal Spread} 
\label{subsec:LongDiff}
A point-like charge deposit will tend to be read with a finite width in the longitudinal direction due to two main effects: the EL gap induced spread and the longitudinal diffusion. The former being due to the light, which ultimately forms the signal, being produced not at a single $z$ point but over the whole distance between the gate and anode. Both effects contribute to the observed signal $z$ sigma in the following way:
\begin{equation}
  \sigma_{t} = \sqrt{{\sigma_{L}}^{2} + {S_{L}}^{2}}
  \label{eq:modeldif}
\end{equation}
where $\sigma_t$ is the sigma in $z$ of the signal, $\sigma_{L}$ is the sigma of the spread induced by longitudinal diffusion, and $S_{L}$ is the EL gap induced longitudinal spread (all units in seconds). The longitudinal diffusion term $\sigma_{L}$ is defined as:
\begin{equation}
  \sigma_{L}^2 = (\frac{2D_{L}}{v_{d}^3})\cdot{}z
  \label{eq:DiffCoef}
\end{equation}
where $D_{L}$ is the longitudinal diffusion coefficient (in units of $[cm^{2}~s^{-1}]$), $v_{d}$ is the electron drift velocity, and z is the drift length.

\subsubsection{EL gap induced Longitudinal Spread $S_{L}$}
\label{subsec:Longabe}
\begin{figure}
  \begin{center}
    \includegraphics[width=0.475\textwidth]{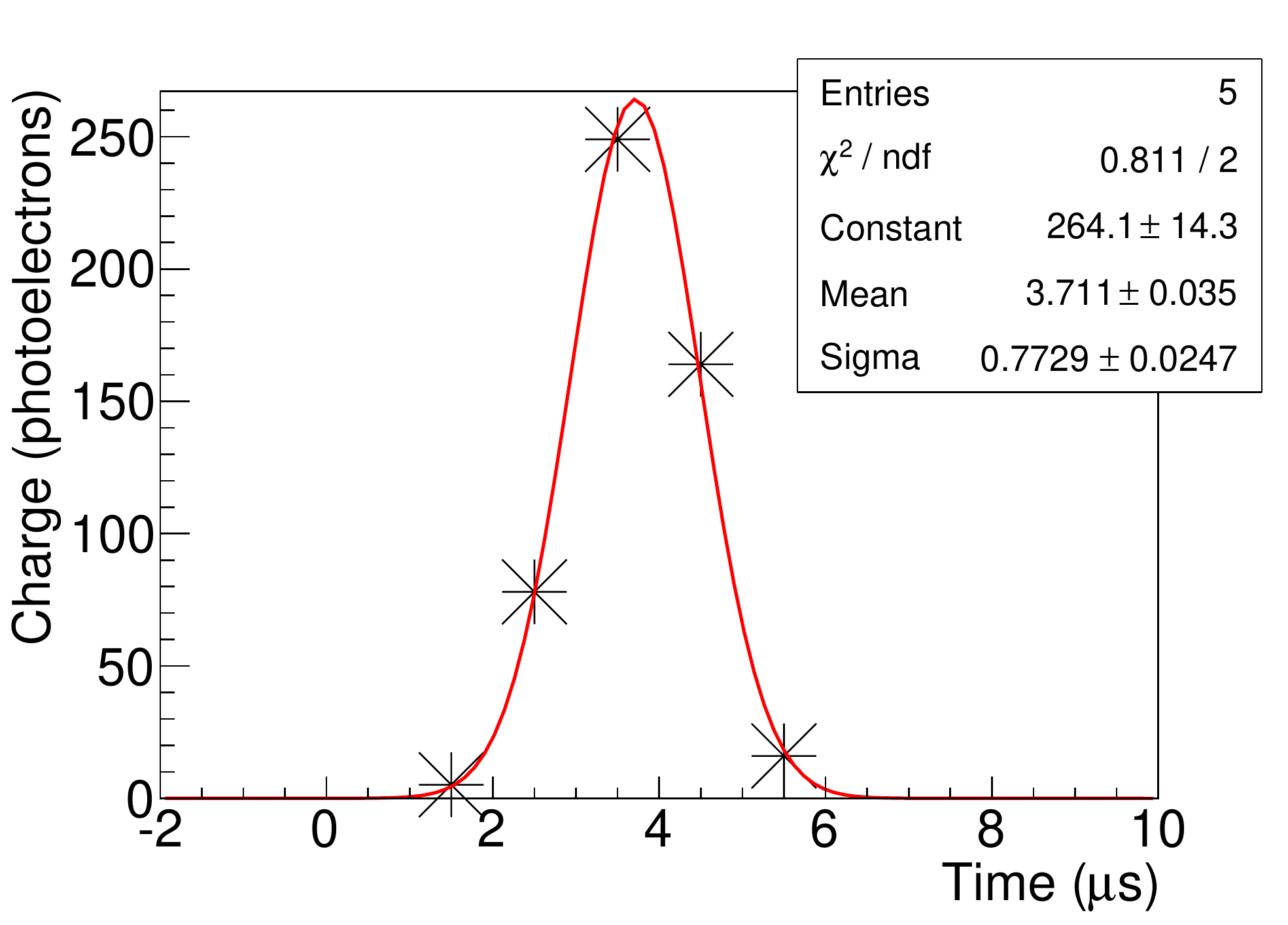} 
    \includegraphics[width=0.505\textwidth]{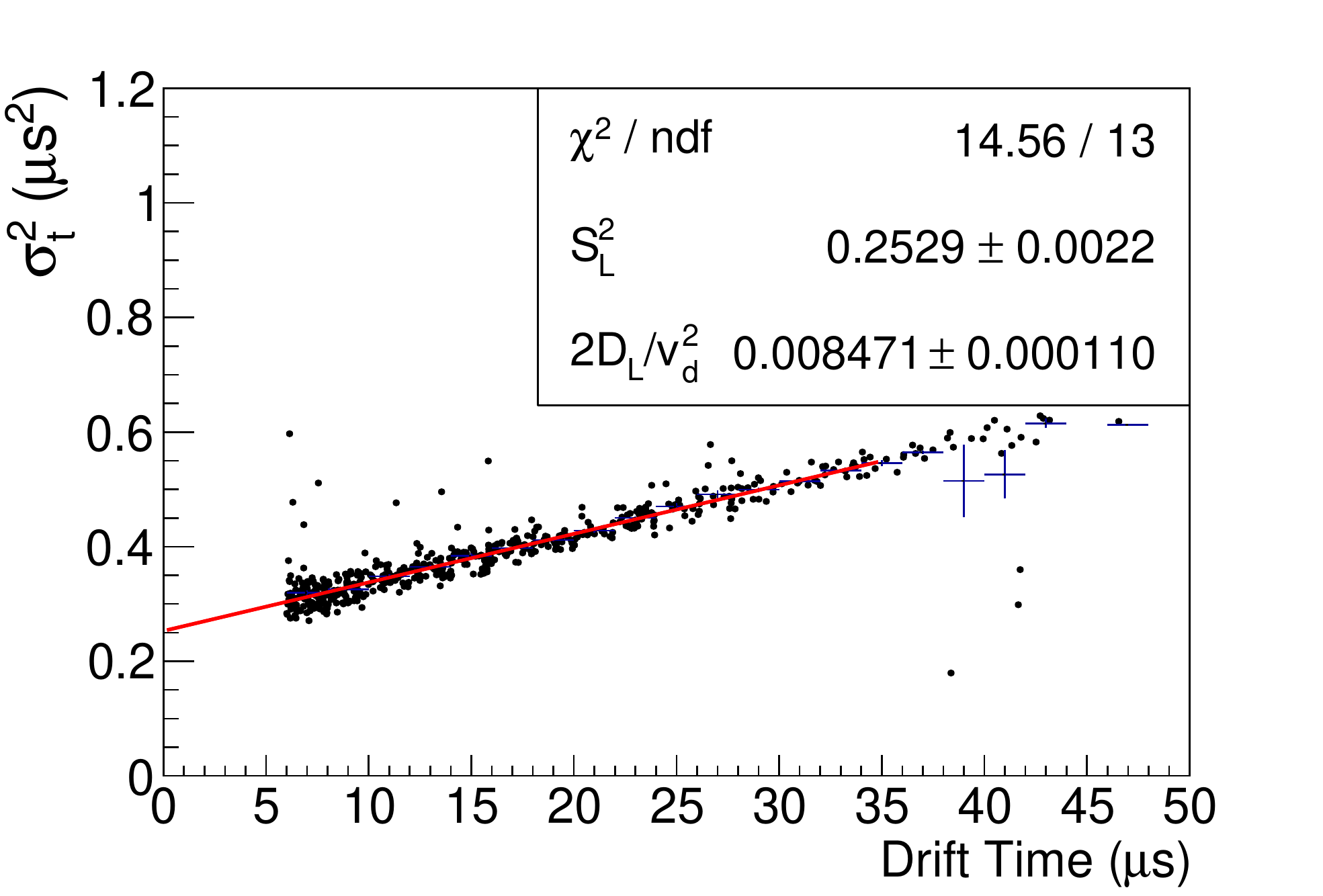} 
    \caption{ Left: Temporal charge distribution of a typical x-ray event and Gaussian fit. Right: Variance of the Gaussian fits versus drift time $t_{d}$, and linear fit to the model obtaining $S_L^2$ as the intercept.}
\label{fig:Longaberra}
\end{center}
\end{figure}

Using \textsl{Configuration 2} data to maximize the number of events at small drift lengths and selecting X-ray deposits using the additional requirement of Gaussian form of the S2 in the $z$ direction, $S_L$ can be studied. Events with drift times ($t_{d}$) less than 50~$\mu$s had their temporal charge distributions fitted (figure \ref{fig:Longaberra}-\emph{left}) with those events not successfully fitted by this model rejected. The number of rejected events is compatible with the number of not X-ray events present in the initial selection made on section~\ref{subsec:Data}.

Using equations \ref{eq:modeldif} and \ref{eq:DiffCoef} it can be seen that there exists a linear relationship between the variance of the signal in $z$ and the drift time with the intercept equal to the square of $S_L$:
\begin{equation}
  \sigma_{t}^{2} = \frac{2D_L}{v_d^2}t_d + {S_{L}}^{2}
  \label{eq:SL}
\end{equation}
where for a set drift field and gas conditions the longitudinal drift coefficient and drift velocity, and hence their ratio, are constant. Therefore, fitting this model to the distribution obtained using the selection mentioned above allows for a determination of $S_L$. For the standard drift field settings (2~kV~cm$^{-1}$~bar) this method yields a value of $S_{L} = 0.50\pm0.05~\mathrm{\mu}$s. This result is slightly different to the value obtained in \cite{Alvarez:2012hu}, where a value of $S_{L} = 0.8~\mathrm{\mu}$s was obtained, presumably because the difference in drift field of the EL region (1~kV~cm$^{-1}$~bar) . At higher field a higher drift velocity reduces the drift time across the EL gap of the electrons. Another implication of this value is the minimum expected duration in $z$ of an event. A sigma of $0.5~\mu$s implies that $\sim$99.7\% of the charge would be contained within 6$\sigma$ and, as such, we would expect a minimum $z$ width of $\sim$3~$\mathrm{\mu}$s, a value which includes the shaping of the electronics.

\subsubsection{Longitudinal Diffusion}
\label{sec:diffL}
Using the same method as in the previous section and extending the allowed drift time, the longitudinal diffusion coefficient can be determined by a fit to the $z$ sigma versus the drift length using the model of equation \ref{eq:modeldif} and the calculated drift velocity from section \ref{subsec:DriftVel} and $S_L$ from section \ref{subsec:Longabe}. 

\begin{figure}
  \begin{center}
    \includegraphics[width=0.49\textwidth]{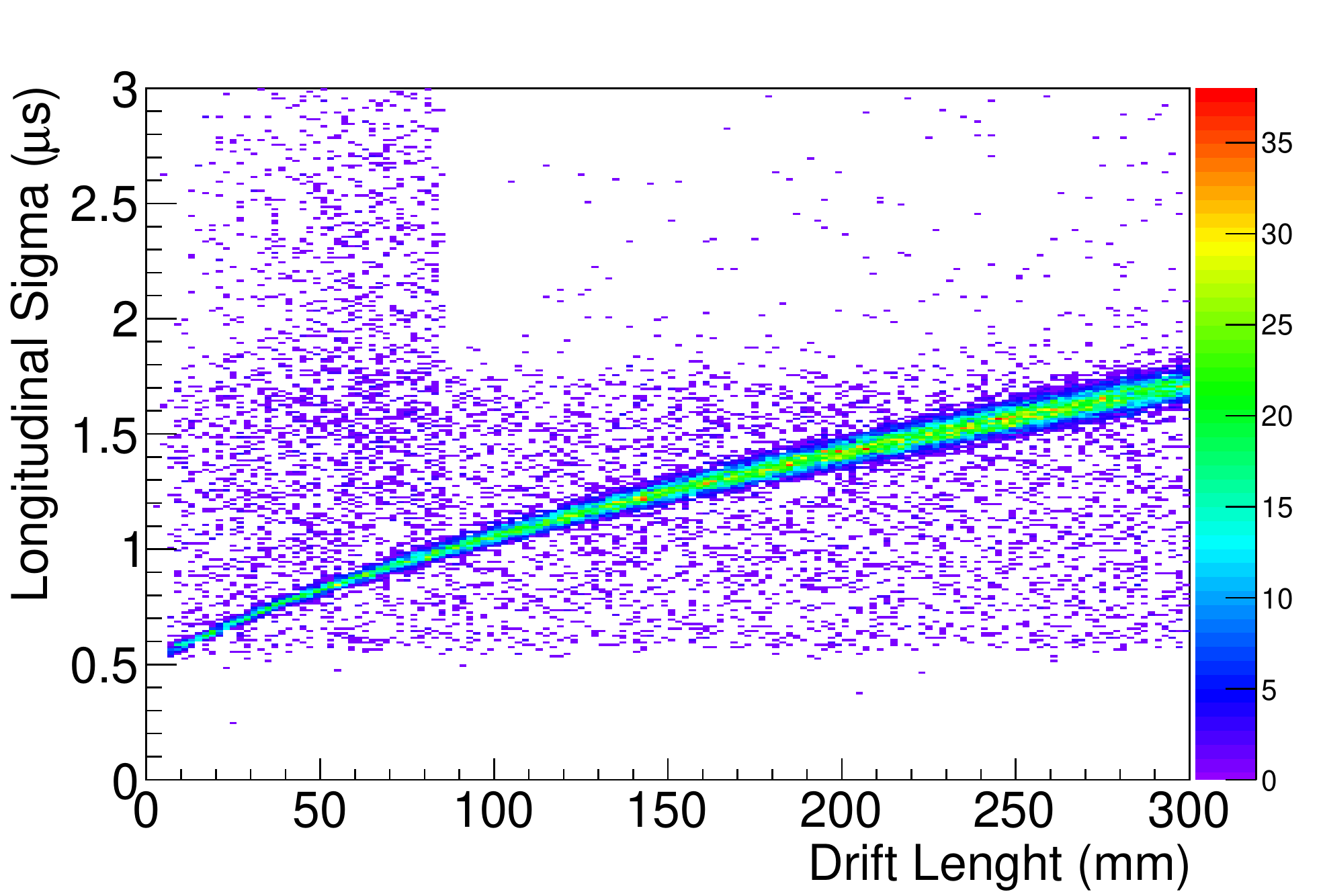} 
    \includegraphics[width=0.49\textwidth]{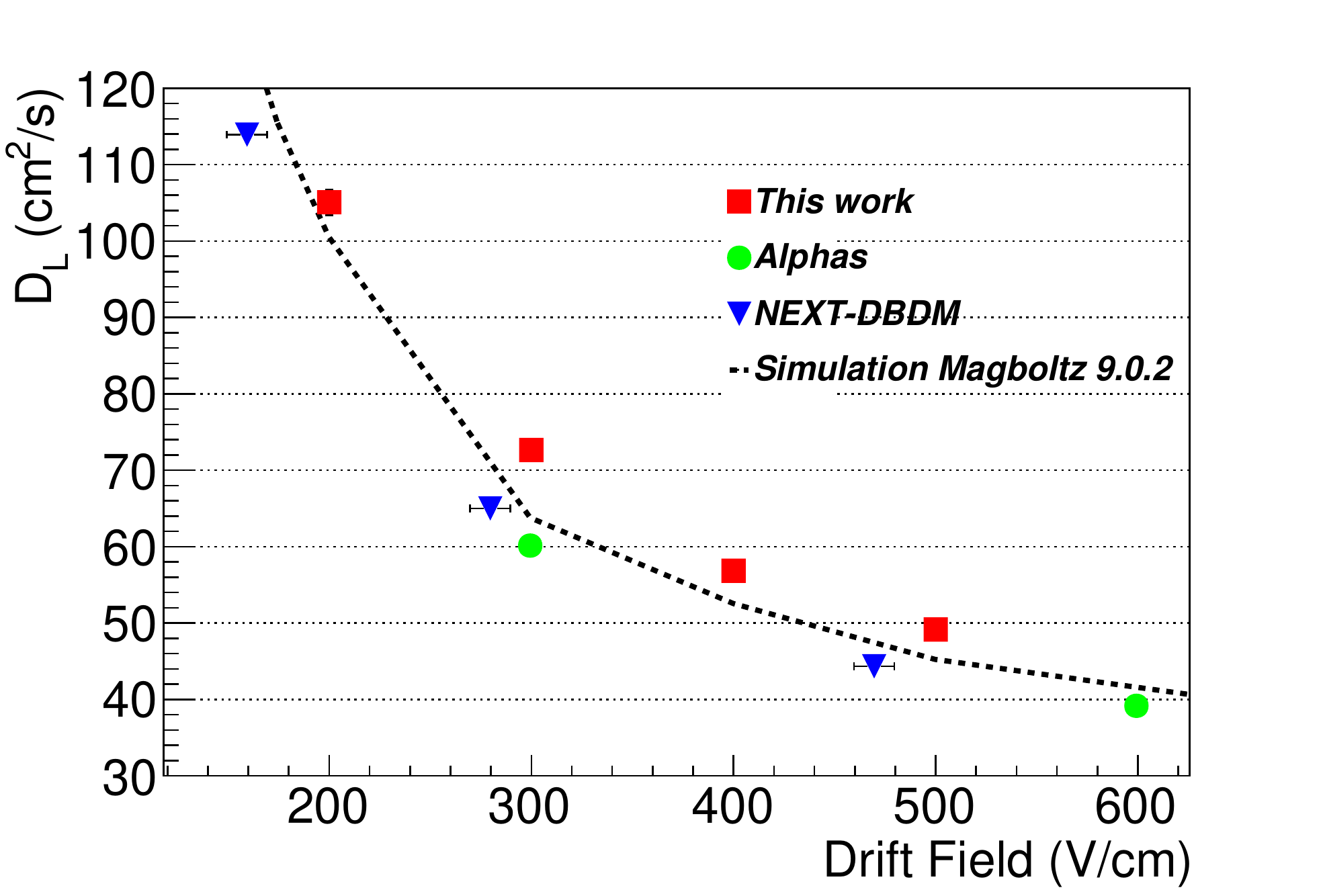} 
    \caption{Left: Longitudinal sigma of the temporal charge distribution fits versus drift length for the 0.5~kV drift field configuration dataset. Right: Longitudinal diffusion coefficient $D_{L}$ as a function of drift field for xenon gas at 10~bar. The red square points are the measured values, green circles correspond to the results of \cite{Alvarez:2012hu}, blue triangles are from \cite{Alvarez:2012kua} while the dashed curve is the prediction for pure xenon at 10~bar from the Magboltz 9.0.2 simulation.}
\label{fig:LongDiff}
\end{center}
\end{figure}

The model is shown in figure \ref{fig:LongDiff}-\emph{left} for a drift field of 0.5 kV~cm$^{-1}$. This coefficient was determined for the same drift field settings used for drift velocity studies of section \ref{subsec:DriftVel} and once again compared to a pure xenon simulation based on version 9.0.2 of Magboltz (figure \ref{fig:LongDiff}-\emph{right}). These results are in slight disagreement with previous data obtained by the \textsl{NEXT Collaboration} as published in \cite{Alvarez:2012kua, Alvarez:2012hu}, where somewhat lower values were extracted. The differences with simulation and previous results are not fully understood at present and could be attributed to difference in gas conditions during data taking.

\subsection{Transverse Spread} 
\label{subsec:Transabe}

The transverse response of the tracking plane to a point-like charge deposition is expected to have a width distribution due to the convolution of the transversal diffusion and the EL gap induced Transverse Spread $S_{T}$. The ionization electrons will diffuse transversely as they drift up to the EL region. Once there, due to the isotropic emission of light, each electron will be seen as the projection of a cone, and therefore, a $K_\alpha$ deposit will be seen as multiple overlapping cones.

\begin{figure}
  \begin{center}
    \includegraphics[width=0.49\textwidth]{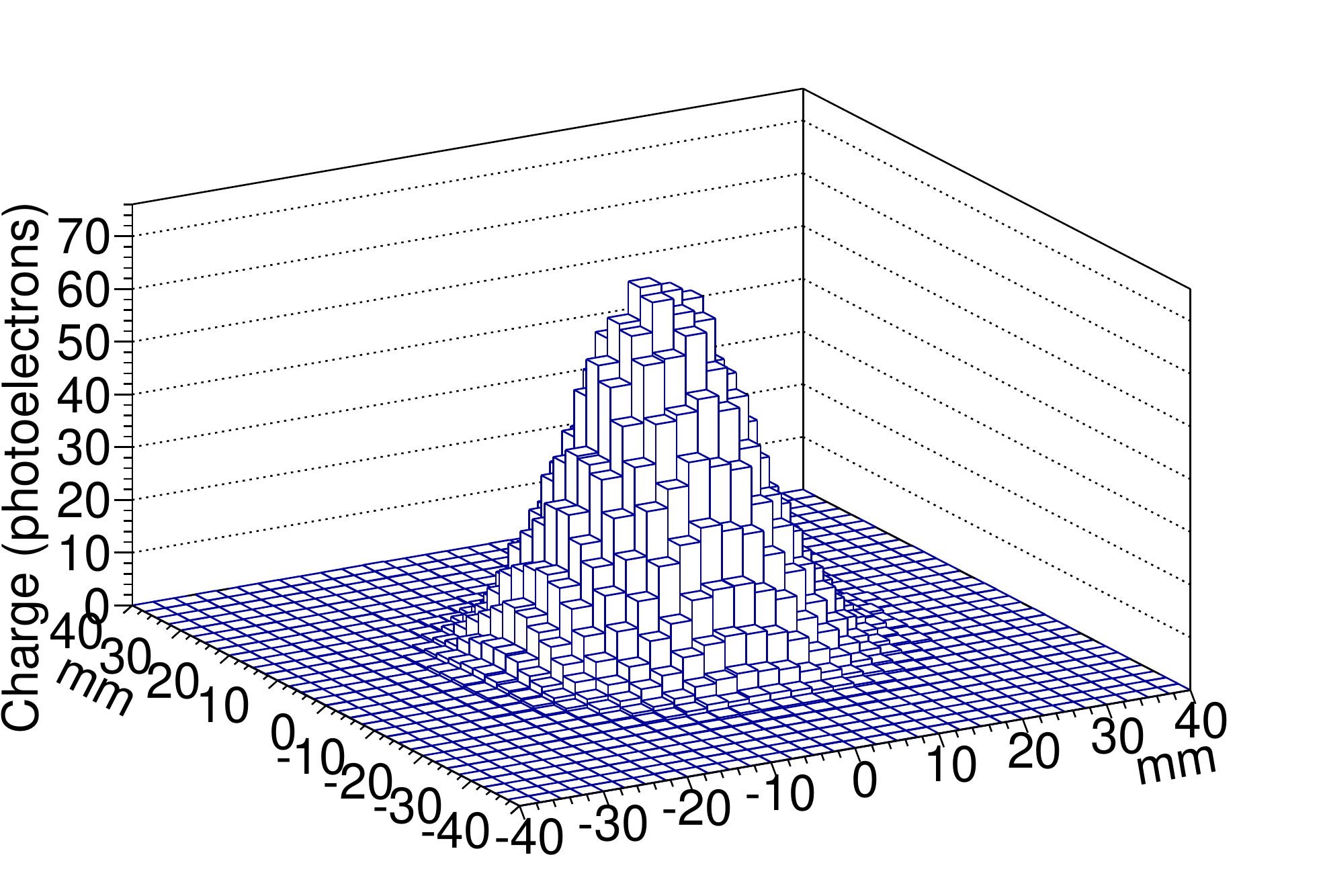} 
    \includegraphics[width=0.49\textwidth]{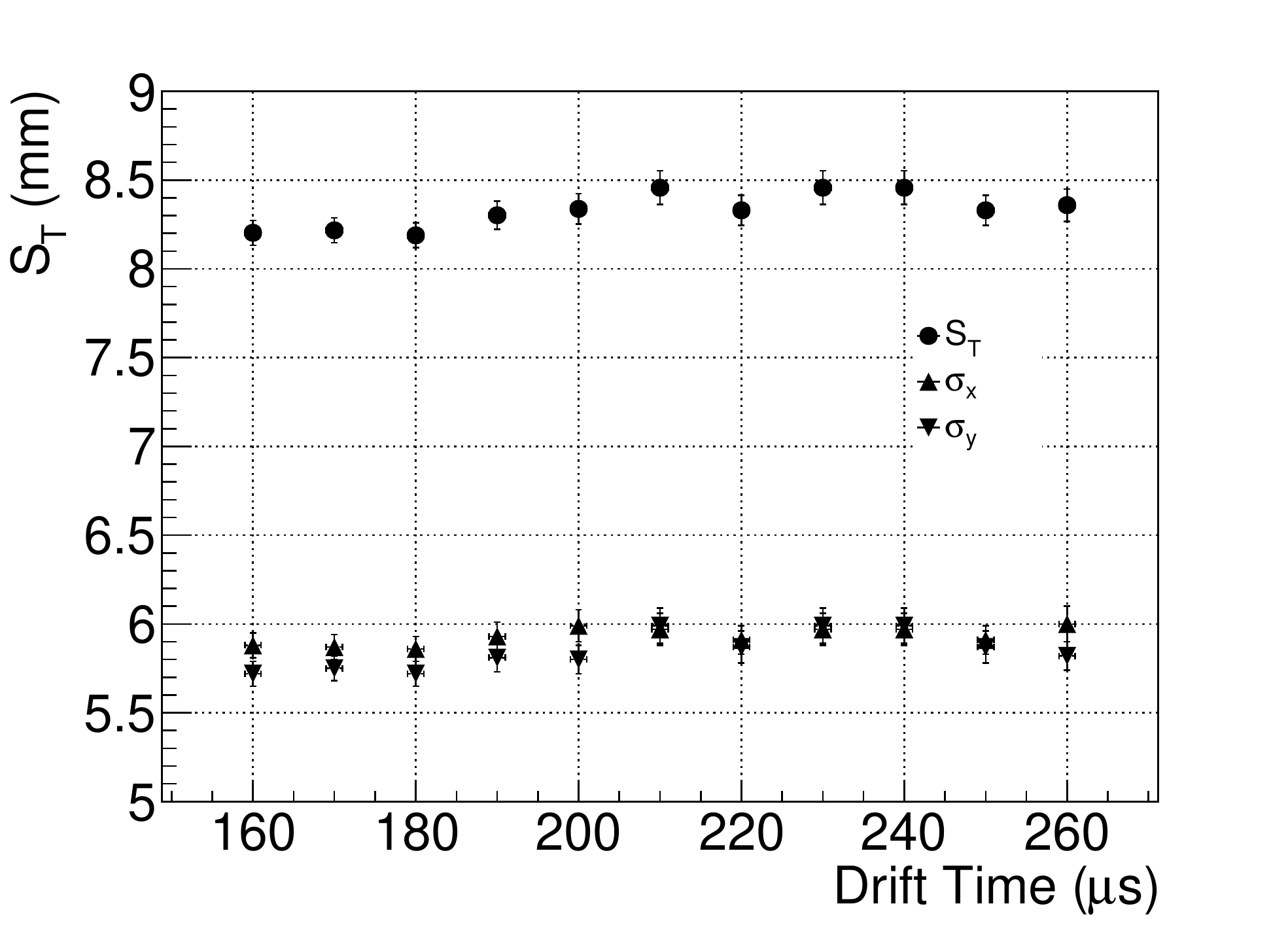} 
    \caption{Left: Average 3D charge distribution of a x-ray event from its barycenter. Right: $S_{T}$,  $\sigma_{x}$ and $\sigma_{y}$ of charge distribution gaussian fit versus Drift Length.}
    \label{fig:transaberr}
  \end{center}
\end{figure}

Using the events selected as X-ray using the criterion mentioned above, a study of the extent of this projection was carried out. Figure \ref{fig:transaberr}-\emph{left} shows the average projection of an event onto the x-y plane with the channel with maximum charge taken as the centre and the charge of the other channels plotted according to their distance from it. A two dimensional Gaussian can be fitted to the distribution to give an estimate of the transverse spread of the charge.

Figure \ref{fig:transaberr}-\emph{right} shows the sigmas of the two dimensional fit as well as their quadratic sum which is the parameter used to define the EL gap induced transverse spread ($S_{T} = \sqrt{{\sigma_{x}}^2 + {\sigma_{y}}^2}$), plotted with drift time. There is no significant trend in the measured values with drift time suggesting that the EL gap distortion dominates the transverse spread of the charge cloud in the detector. The pitch of the SiPM channels in NEXT-DEMO is 10~mm, $S_T$ of the order of 8~mm suggests that little would be gained by increasing sensor density.
%

\section{Energy Resolution} 
\label{sec:EnergyResolution}
One of the most important goals of NEXT-DEMO is to prove that the energy of electron tracks can be reconstructed accurately and that the resolution calculated for these tracks can be extrapolated to $\Qbb$. While the energy resolution of the raw data, only considering the online trigger, is already good, there are a number of correctible detector effects which can be understood and equalised optimizing the energy resolution of the detector. Among these effects are attachment during drift which causes a drift distance dependant energy measurement and inhomogeneities in light production and reflection due to the grids and light tube. As described in section \ref{sec:DetRes}, these effects can be understood using the $K_\alpha$ X-ray events and used to equalise the detectors response.

\subsection{Attachment Correction} 
\label{subsec:AttCorr}
\begin{figure}
  \begin{center}
    \includegraphics[width=0.7\textwidth]{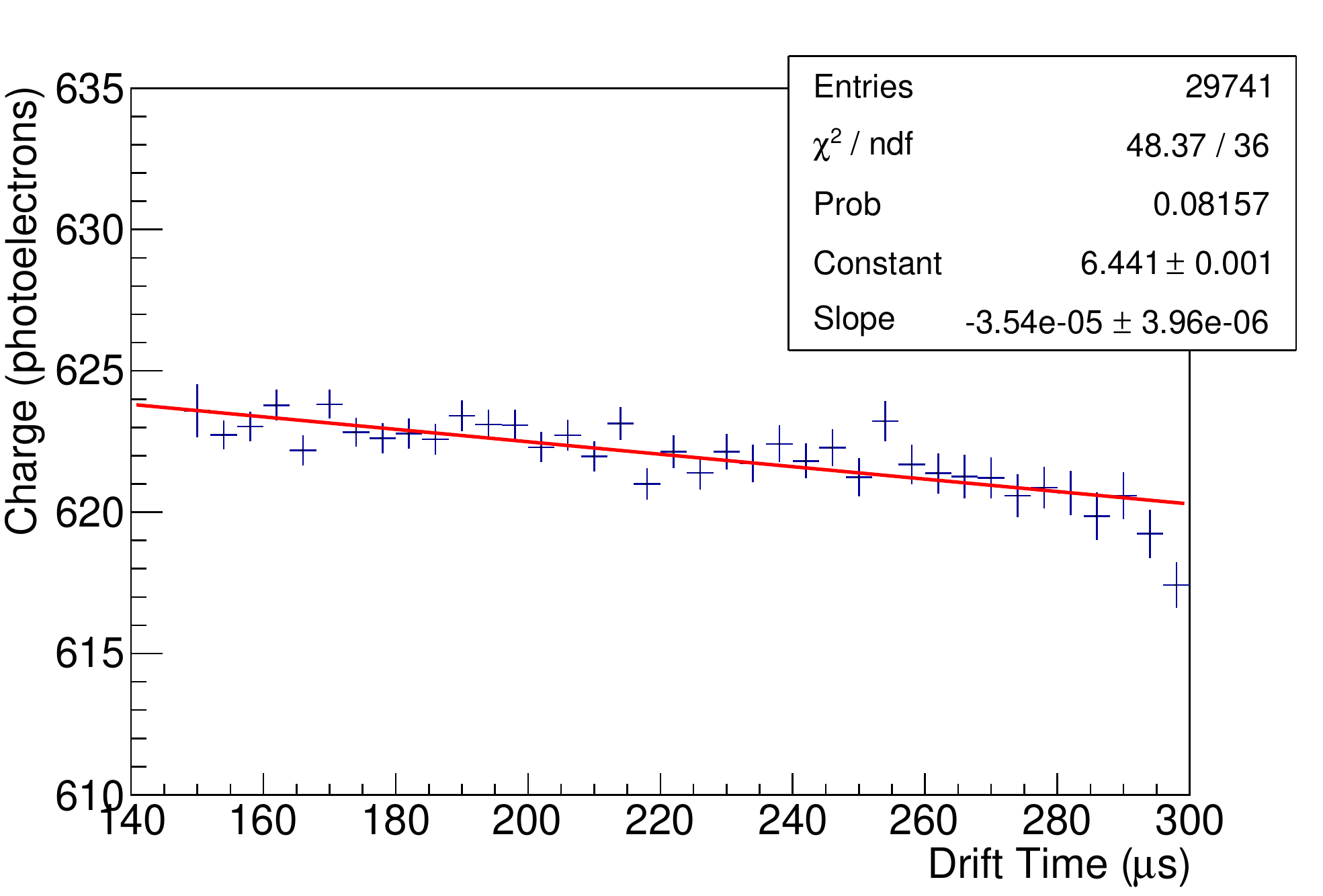} 
    \caption{Mean charge of X-ray events versus drift time. Exponential fit is made and correction factor extracted.}
    \label{fig:Zplot}
  \end{center}
\end{figure}

During the charge cloud's drift towards the anode, a fraction of the charge is lost due to attachment. Attachment is limited by the constant circulation and cleaning of the gas through hot getters but there remains a small, observable effect. Using the $K_\alpha$ peak in the selected data and plotting its charge with drift time this decay can be seen (figure \ref{fig:Zplot}). Modelling this decay as an exponential, a value for the mean electron lifetime can be extracted. Using this value the loss of charge due to attachment can be effectively corrected.
In all datasets considered  in this paper the decay constant was of order $-3.5\times10^{-5}~\mu$s$^{-1}$, equivalent to a mean electron lifetime of $\sim$28~ms, far larger than the maximum drift time of the TPC.
This value represents an improvement over the values achieved in other gaseous xenon detectors \cite{Dobi2011215} .

\subsection{Calculation of energy weighted sum} 
\label{subsec:Extrapol}
The correction factor determined above together with those obtained in section \ref{subsubsec:XYCorr} can, in principle, be used as a model for the correction of the energy of any type of interaction in the TPC. While an extended event will require consideration in time slices, the photoelectric events of \NA\ are still small enough that the X-ray model is a good one.

The corrected event energy is calculated as the weighted sum of the contributions from each individual PMT:
\begin{equation}
  Q_{tot} = S_{0}(x,y) \cdot{} \sum_{i} q_{i} \cdot{} w_{i}(x,y) \cdot{} f_i(x,y)
  \label{eq:Esum}
\end{equation}
where $q_{i}$ is the charge recorded by PMT $i$, $w_{i}(x,y)$ its weight for the reconstructed $(x,y)$ position -- here, the inverse of the variance of its response to $K_\alpha$ X-rays as described in section \ref{subsubsec:QE} -- and $f_{i}(x,y)$ is the geometrical correction factor for PMT $i$ in for the reconstructed $(x,y)$ position as explained in Section \ref{subsubsec:XYCorr}. The term $S_{0}(x,y)$ is an overall conversion factor from photoelectrons to energy. Using this weighted energy estimator as opposed to the basic mean value used in previous publications (for example \cite{Alvarez:2013gxa}) improves the determination of the energy by taking into account inhomogeneities in the response of the PMTs.

The residuals of the Delaunay interpolation produce a slight difference in the energy scale according to the $(x,y)$ bin in which an event falls. This is accounted for by calculating a scaling factor per bin, $S_{0}(x,y)$. These factors are calculated by fitting a straight line, in each bin, to the known energy of well defined peaks (the $K_\alpha$ and $K_\beta$ peaks at 29.7~keV and 33.8~keV respectively) and the photoelectric peak and its escape peak (in the case of \NA\ at 511~keV and $\sim$481~keV).

\begin{figure}
  \begin{center}
    \includegraphics[width=0.7\textwidth]{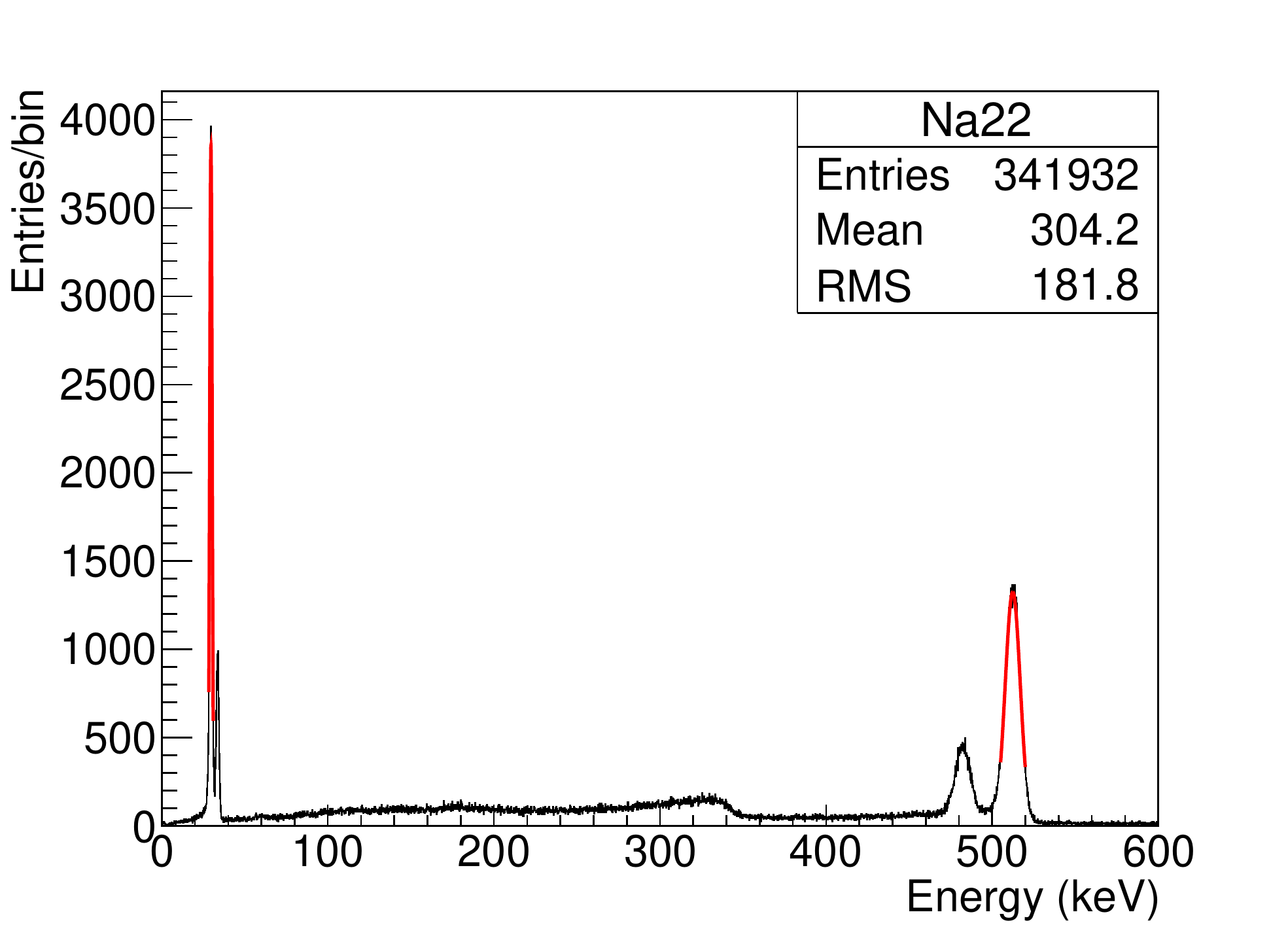} 
    \includegraphics[width=0.49\textwidth]{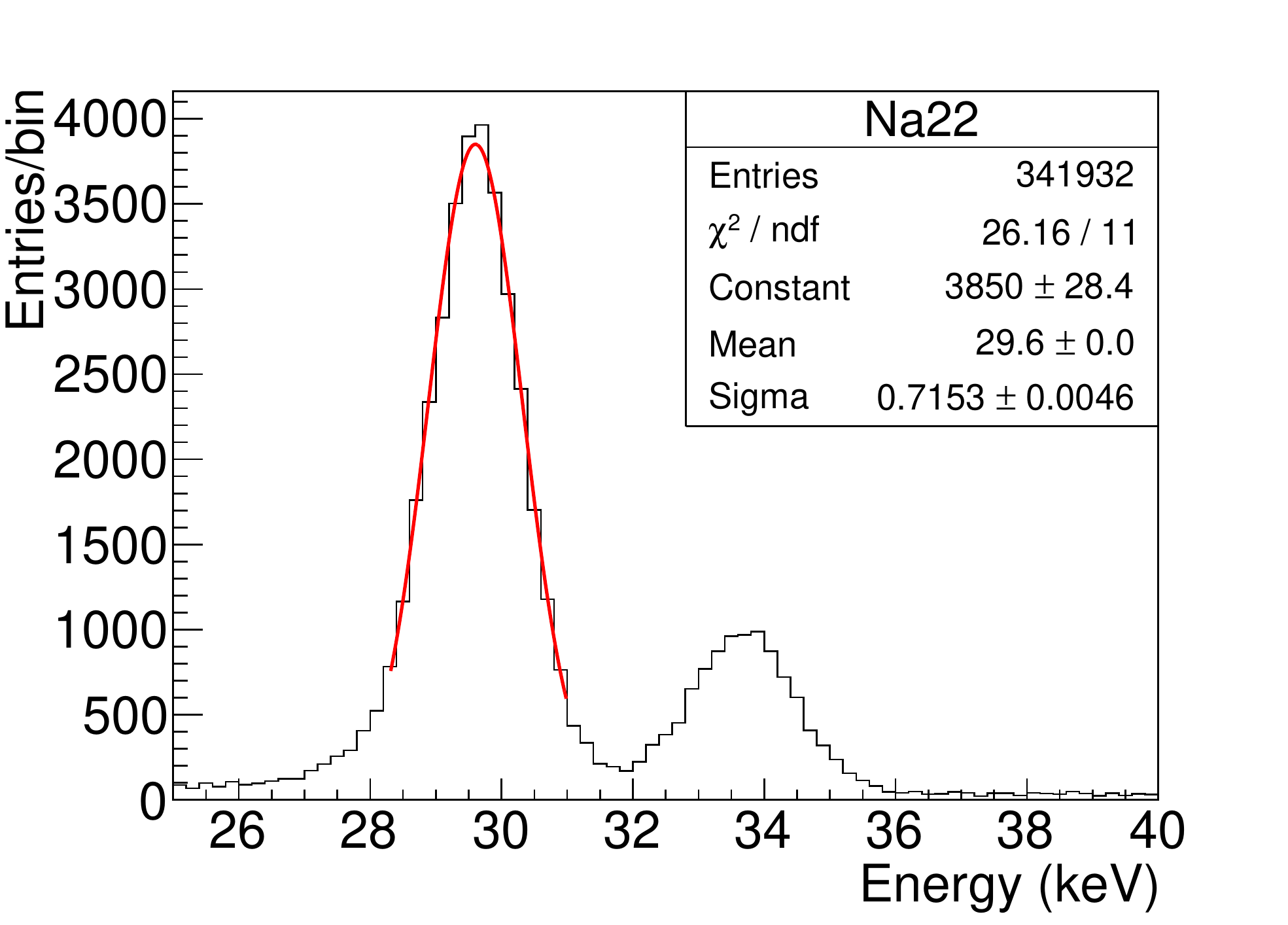} 
    \includegraphics[width=0.49\textwidth]{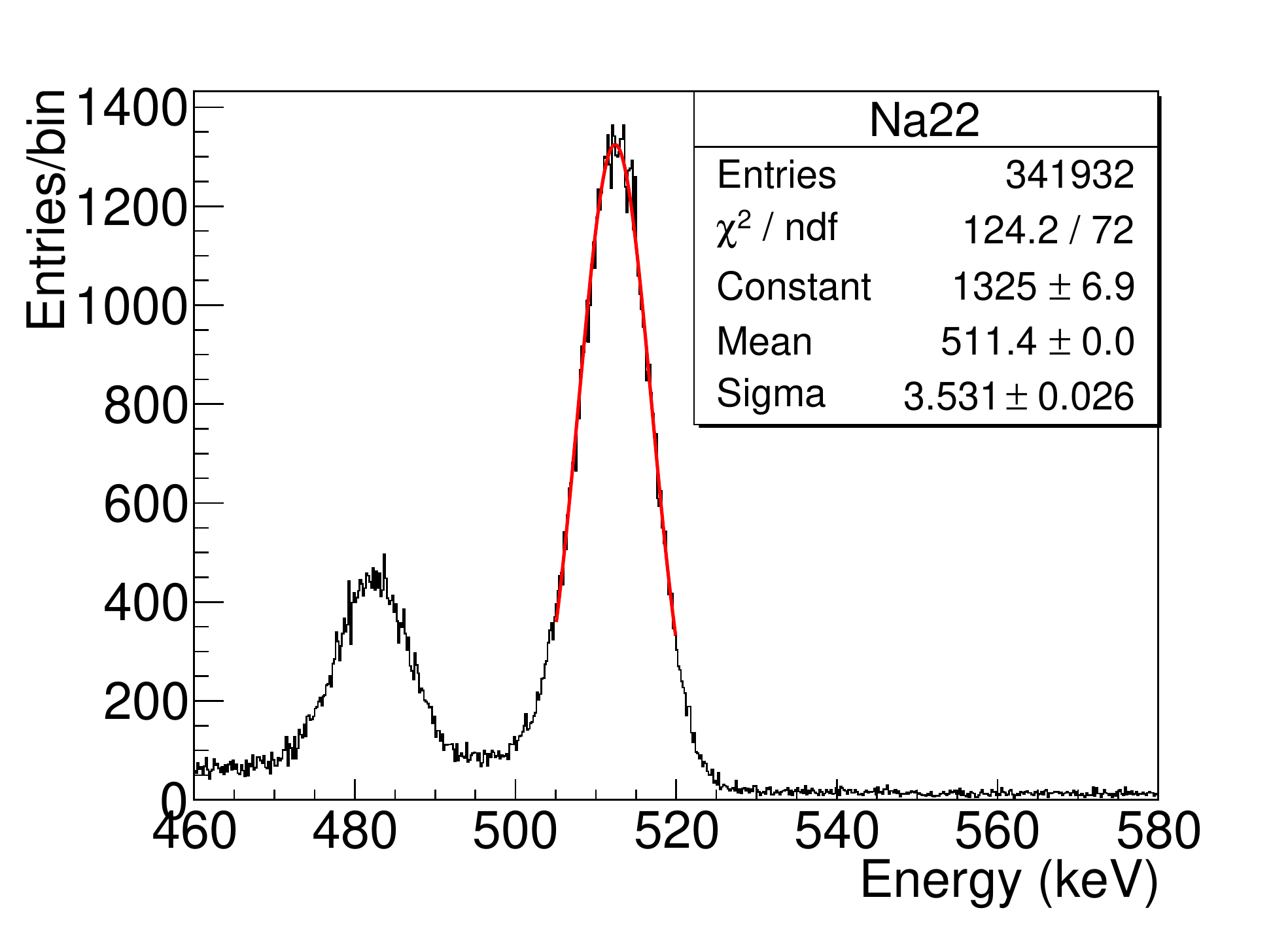}
    \caption{Top: \NA\ spectrum after all corrections. Bottom-Left: Zoom of X-ray peak region and gaussian fit of the $K_\alpha$ peak. Bottom-Right: Zoom of the escape and photoelectric peaks and gaussian fit to the Photoelectric.}
\label{fig:SpecER}
\end{center}
\end{figure}

Applying all the corrections described, the resultant \NA\ spectrum is that shown in figure \ref{fig:SpecER}-\emph{top} with energy resolution for the $K_\alpha$ peak of ($5.691\pm0.003$)\% FWHM (detail shown in figure \ref{fig:SpecER}-\emph{bottom left}). The \NA\ photopeak has a resolution of ($1.62\pm0.01$)\% FWHM. Independently extrapolating these two values to the \XE\ \Qbb\ assuming the dominance of photon shot noise Poisson statistics results in predicted energy resolution at \Qbb\ of 0.6256\% FWHM extrapolating from the $K_\alpha$ peak and 0.7353\% FWHM from the photopeak. The discrepancy in the two values can be attributed to the gradual breakdown of the model of X-ray response as a global correction as extended events become more probable. 
%

\section{Summary} 
\label{sec:Summary}
NEXT-DEMO data has been used to present the flexibility of xenon X-ray events both as a means to understand the fundamental properties of the TPC and as a model for the equalisation of detector energetic response.

The drift velocity and diffusion of the TPC have been determined using these events, which are abundant in any type of data taking. These properties can be used to monitor the gas quality of the detector and to facilitate the union of datasets with slightly different conditions. Moreover, they can be used to understand the effect of the EL gap on the observed signals.

The same events have also been used to understand inhomogeneities in the detector response allowing for a normalisation of effects due to uneven deposition of wavelength shifter and asymmetries in the form of the light tube. This model has been used to calculate a corrected weighted sum of the observed energy resulting in \NA\ photopeak energy resolution of 1.62\% and a predicted \Qbb\ resolution of as little as 0.63\%. These values represent a slight improvement on previously published results \cite{Alvarez:2013gxa}.

Future work will involve the generalisation of the X-ray model so it can be used to correct more extended events where temporal slicing and in slice deposit clustering are required to precisely equalise the energetic response of the TPC.

\acknowledgments
This work was supported by the following agencies and institutions: the European Research Council under the Advanced Grant 339787-NEXT; the Ministerio de Econom\'ia y Competitividad of Spain under grants CONSOLIDER-Ingenio 2010 CSD2008-0037 (CUP), FPA2009-13697-C04 and FIS2012-37947-C04; the Director, Office of Science, Office of Basic Energy Sciences, of the US Department of Energy under contract no.\ DE-AC02-05CH11231; and the Portuguese FCT and FEDER through the program COMPETE, project PTDC/FIS/103860/2008.

\bibliographystyle{JHEP}
\bibliography{references}

\end{document}